%% file: ms.tex
\documentclass[sigconf]{acmart}
\pdfoutput=1

\newcommand{\ignore}[1]{}
\usepackage[normalem]{ulem}

\usepackage[hyphenbreaks]{breakurl}
\usepackage{hyperref}
\usepackage{array}
\usepackage{multirow}
\usepackage{subfig}
\usepackage{arydshln}
\usepackage{siunitx}
\usepackage{enumitem}
\usepackage{epigraph}

\newcommand\javier[1]{\noindent{\color{cyan} {\bf \fbox{Javier}} {\it#1}}}
\newcommand\dionisis[1]{\noindent{\color{cyan} {\bf \fbox{Dionisis}} {\it#1}}}
\newcommand\alex[1]{\noindent{\color{red} {\bf \fbox{Alex}} {\it#1}}}
\newcommand\dmitrii[1]{\noindent{\color{brown} {\bf \fbox{Dmitrii}} {\it#1}}}

\newcommand\mjs[1]{\noindent{\color{blue} {\bf \fbox{MJS}} {\it#1}}}

\title{{Storage-Class Memory Hierarchies for Servers}}
\title{{Enabling Storage Class Memory as a DRAM Replacement \mbox{for Datacenter Services}}}
\title{Design Guidelines for High-Performance SCM Hierarchies}

\begin{document}

	\author{Dmitrii Ustiugov}
	\affiliation{\institution{EcoCloud, EPFL}}
	\email{dmitrii.ustiugov@epfl.ch}

	\author{Alexandros Daglis}
	\affiliation{\institution{EcoCloud, EPFL}}
	\email{alexandros.daglis@epfl.ch}
	
	\author{Javier Picorel}
	\affiliation{\institution{Huawei Technologies}}
	\email{javier.picorel@huawei.com}
	
	\author{Mark Sutherland}
	\affiliation{\institution{EcoCloud, EPFL}}
	\email{mark.sutherland@epfl.ch}

	\author{Edouard Bugnion}
	\affiliation{\institution{EcoCloud, EPFL}}
	\email{edouard.bugnion@epfl.ch}

	\author{Babak Falsafi}
	\affiliation{\institution{EcoCloud, EPFL}}
	\email{babak.falsafi@epfl.ch}
	
	\author{Dionisios Pnevmatikatos}
	\affiliation{\institution{FORTH-ICS \& ECE-TUC}}
	\email{pnevmati@ics.forth.gr}

\begin{abstract}
    With emerging storage-class memory (SCM) nearing commercialization, there is evidence that it will deliver the much-anticipated high density and access latencies within only a few factors of DRAM. Nevertheless, the latency-sensitive nature of memory-resident services makes seamless integration of SCM in servers questionable. In this paper, we ask the question of how best to introduce SCM for such servers to improve overall performance/cost over existing DRAM-only architectures. We first show that even with the most optimistic latency projections for SCM, the higher memory access latency results in prohibitive performance degradation. However, we find that deployment of a modestly sized high-bandwidth 3D stacked DRAM cache makes the performance of an SCM-mostly memory system competitive. The high degree of spatial locality that memory-resident services exhibit not only simplifies the DRAM cache's design as page-based, but also enables the amortization of increased SCM access latencies and the mitigation of SCM's read/write latency disparity. 
    
    We identify the set of memory hierarchy design parameters that plays a key role in the performance and cost of a memory system combining an SCM technology and a 3D stacked DRAM cache. We then introduce a methodology to drive provisioning for each of these design parameters under a target performance/cost goal. Finally, we use our methodology to derive concrete results for specific SCM technologies. With PCM as a case study, we show that a two bits/cell technology hits the performance/cost sweet spot, reducing the memory subsystem cost by 40\% while keeping performance within 3\% of the best performing DRAM-only system, whereas single-level and triple-level cell organizations are impractical for use as memory replacements.
    

\end{abstract}


\copyrightyear{2018} 
\acmYear{2018} 
\setcopyright{acmcopyright}
\acmConference[MEMSYS]{The International Symposium on Memory Systems}{October 1--4, 2018}{Old Town Alexandria, VA, USA}
\acmBooktitle{The International Symposium on Memory Systems (MEMSYS), October 1--4, 2018, Old Town Alexandria, VA, USA}
\acmPrice{15.00}
\acmDOI{10.1145/3240302.3240310}
\acmISBN{978-1-4503-6475-1/18/10}


\maketitle
\date{}

\thispagestyle{empty}
\pagestyle{plain}
\sloppy


\input{intro}

\input{background}

\input{server_workloads}
\input{scm_space}
\input{methodology}

\input{evaluation}

\input{discussion}

\input{related}
\input{conclusion}
\input{ack}


\bibliographystyle{ACM-Reference-Format}
\bibliography{tools/gen-abbrev,dblp,ref}

\end{document}

%% file: intro.tex
\section{Introduction}
\label{sec:intro}

For almost 50 years, DRAM has served as the universal standard for memory in mainframes, laptops, and datacenters.
Particularly in the datacenter, we are entering a new age where memory will no longer exclusively be comprised of DRAM.
Although the interactive nature of online services will continue to dictate that hot data must be kept DRAM resident, capacity and cost limitations have begun to pressure datacenter operators to investigate emerging technologies to replace it.
Future servers will undoubtedly retain some DRAM for performance, while shifting to denser main memories to hold vast datasets.

As in-memory datasets have continued growing exponentially~\cite{amazon:ec2,keeton:memory}, memory architects have been unable to provide products with sufficient capacity, obstructed by fundamental limitations on channels per packaged IC as well as intra-channel signal integrity.
With the pressure squarely on DRAM manufacturers to deliver DIMMs with ever-increasing capacities, memory has begun to form a significant fraction of server acquisition cost\footnote{With a commodity Xeon E5-2660 v4 CPU and 256GB of DRAM, the memory represents up to 40\% of the server's total acquisition cost \cite{dramexchange,opencompute}.}, as high density components command higher margins and therefore prices.
The synthesis of these two trends has led to a concerted effort to provision memory systems with reduced cost per bit, markedly reducing expenditure for large volume deployments.

Emerging storage-class memory (SCM) technologies are a prime candidate to serve as the next generation of main memory, as they boast approximately an order of magnitude greater density than DRAM at a lower cost per bit~\cite{intel:3dxpoint,wd:reram,computerworld:3dxpoint,zhou:hnvm}.
These traits come at the price of elevated access latency compared to DRAM, creating new challenges for systems designers as memory latency is a critical factor in datacenter application performance~\cite{kanev:profiling}.
Given that typical SCM latencies are 4--100$\times$ greater than DRAM~\cite{suzuki:nonvolatile,wong:stanford}, and that SCM devices often have write latencies 2--10$\times$ longer than reads, na\"ively and completely replacing DRAM with SCM is an unacceptable compromise for datacenter operators.

In addition to the self-evident latency problem, SCM devices come in many flavors, with an inverse relationship between latency and density---typically, denser devices are cheaper but slower.
Ideally, one would like to use the cheapest, highest capacity devices, but their latencies will degrade application performance the most.
In this paper, we identify an opportunity to drastically increase server memory capacity at lower acquisition cost with the use of denser SCM, judiciously retaining a small amount of DRAM for performance reasons.


To realize that opportunity, we replace conventional DRAM with a two-tier hierarchy, combining high-density SCM with a modestly sized 3D stacked DRAM cache; the former component offers cheap capacity and reduced cost, while the latter preserves the low latency and high bandwidth required to ensure interactivity for online services.
The structure of our hierarchy is informed by the insight that SCM's longer access latencies can be amortized with large transfers (e.g., reading or writing KBs of data); therefore we show that the stacked DRAM cache is best organized with page-size blocks whose sizes match with the SCM's row buffers.
Using a 3D stacked DRAM cache aggregates the application's fine-grained accesses while a block is cache-resident, creating bulk transfers to and from the SCM and therefore amortizing its latency. 
Figure~\ref{fig:intro_hierarchy}a is a block diagram of our proposed SCM hierarchy.
We show that by carefully provisioning both levels, datacenter operators can reduce the cost of memory by 40\%, while maintaining performance within 3\% of the best performing DRAM-only system.

The plurality of available SCM and 3D stacked DRAM devices complicates designing such a memory hierarchy.
System designers will be faced with choices pertaining to SCM latency, capacity, memory technology, and form factor; on the 3D stacked DRAM cache side, its high design and integration costs~\cite{dong:fabrication} diminish the returns in cost/bit attained by replacing DRAM with SCM, requiring its parameters (e.g., capacity and block size) to be judiciously chosen.
Hence, the design space for two-level hierarchies is vast, spanning large performance and cost ranges.

To guide architects through this design space, we devise an exploration methodology for any DIMM-packaged SCM technology.
Our methodology operates as follows: based on our insight that SCM latencies can be amortized with bulk transfers, we identify the key design parameter for any SCM device as its \textit{row buffer size}, which sets the upper bound on the transfer size to/from the data array.
For that given row buffer size, we bound the maximum acceptable SCM \textit{read} and \textit{write latencies} (implicitly the minimum SCM device cost), that preserve application performance.
We frame these three parameters as a volume, where all SCM devices within the volume are acceptable choices (an example is shown in Figure~\ref{fig:intro_hierarchy}b).
Our methodology helps designers pinpoint the most cost-effective hierarchies that still meet application performance targets.\\

\input{floats/float-intro}

To summarize, our main contributions are the following:
\begin{itemize}[leftmargin=*]
    \item Analyzing emerging DIMM-packaged SCM devices, and conclusively showing that even the fastest among them cannot directly replace DRAM due to their access latencies. 
    \item Proposing an SCM-based memory hierarchy whose performance is within 10\% of the best DRAM-only system. The hierarchy consists of SCM main memory and a modestly sized DRAM cache. The DRAM cache amortizes the SCM's elevated latency by aggregating many fine-grained accesses into large bulk transfers; furthermore, we show that the DRAM cache must be 3D stacked to cope with high bandwidth demands of today's servers.
    \item Identifying the set of key design parameters for hybrid SCM-DRAM hierarchies, then devising a methodology to prune the vast design space and identify SCM device configurations that offer the highest density, while maintaining performance within 10\% of the best DRAM-only system. Interestingly, we find that the right combination of SCM row buffer and DRAM cache sizing obviates all performance concerns related to the read/write latency disparity inherent in SCM technologies.

    \item Conducting a case study on emerging phase-change memory (PCM) devices, and demonstrating that 2-bit cell organizations (MLC) represent the only cost-effective choice, while both 1-bit and 3-bit cells fail to improve the server's performance/cost ratio. MLC-based memory hierarchies are up to 1.8$\times$ more cost-effective than their DRAM-based counterparts and deliver comparable performance.

\end{itemize}

The rest of the paper is organized as follows: 
First, we describe emerging SCM technologies in \S\ref{sec:background} and motivate our insight to amortize SCM latency with bulk transfers. 
We then analyze server workloads in \S\ref{sec:workloads} to show that na\"ively replacing SCM with DRAM is impractical without the use of a 3D stacked page-based DRAM cache, and identify the cache's critical parameters.
As said cache inflates the system's cost, we introduce a design space exploration methodology in \S\ref{sec:approaches} to drive the search for the most cost-effective system.
Based on our evaluation methodology presented in \S\ref{sec:method}, we provide sample parameters for the SCM design space for servers, and perform a case study with emerging phase-change memory in \S\ref{sec:eval}. 
We discuss additional relevant aspects of hybrid memory hierarchy design in \S\ref{sec:disc} and related work in \S\ref{sec:related}. Finally, \S\ref{sec:conclusion} concludes.

%% file: floats/float-intro.tex
\begin{figure}[t]
\centering
        \includegraphics[width=\columnwidth]{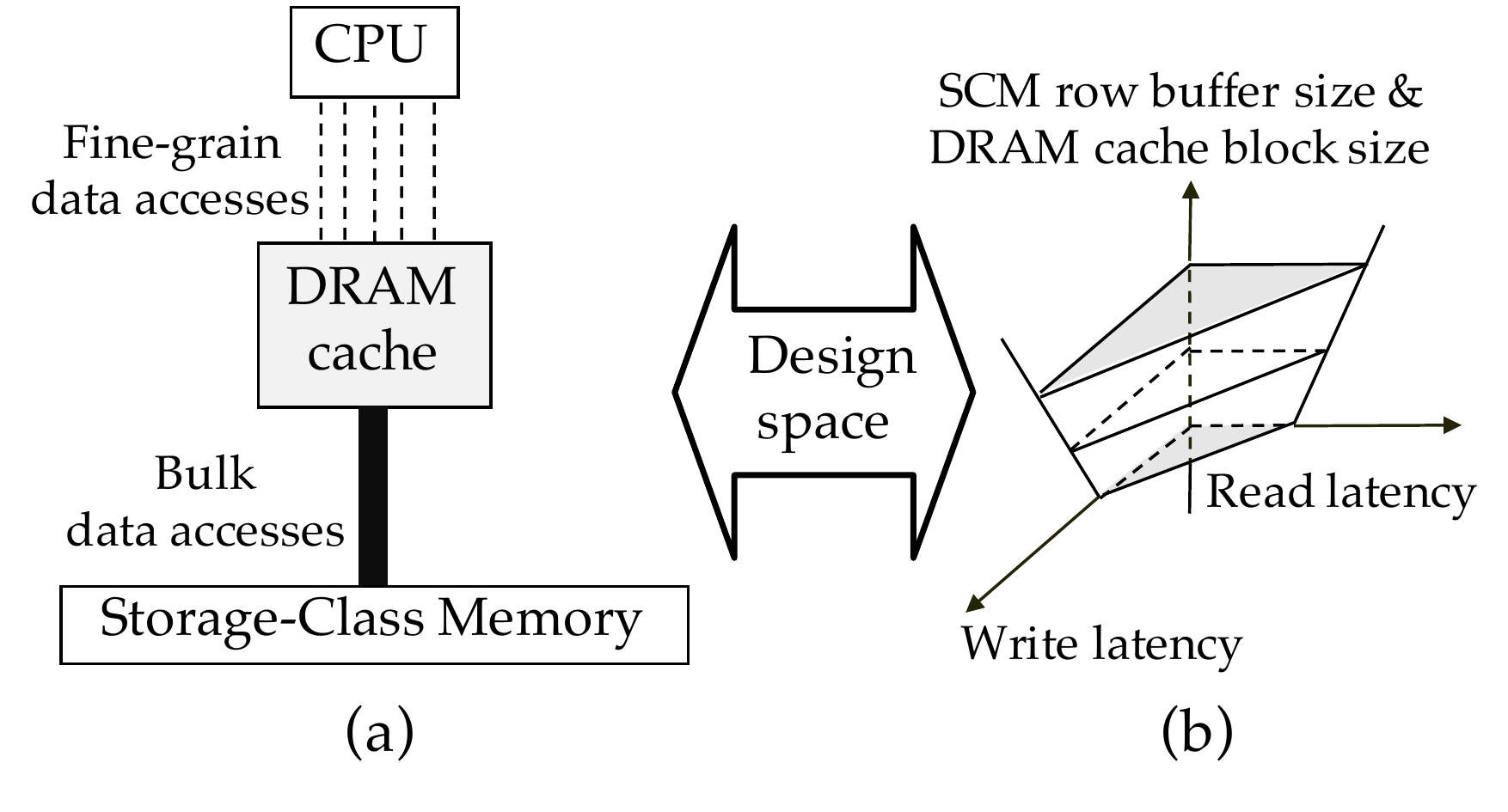}
\caption{SCM hierarchy and design space exploration methodology.}
\label{fig:intro_hierarchy}
\vspace*{-0.6cm}
\end{figure}

%% file: background.tex
\section{SCM Background}
\label{sec:background}

Storage-class memory (SCM) is a term that encapsulates a class of emerging and much-anticipated technologies that are expected to penetrate the computing market in the following decade \cite{intel:3dxpoint,mpr:xpoint,wd:reram}. 
Being slower and denser than DRAM, but faster than Flash while retaining persistence, it cannot be strictly classified as either memory or storage but has characteristics of both. 
While the first SCM products were marketed as faster block-based storage devices, memory vendors have recently launched SCMs intended for drop-in compatibility with commodity DRAM infrastructure; these SCMs are packaged in a dual in-line memory module (DIMM) form factor and use the conventional DDR interface~\cite{computerworld:3dxpoint}. 
Products of the latter flavor will be disruptive for modern servers, as their increased densities will translate into a commensurate reduction in memory provisioning cost.

Although designing SCM DIMMs for compatibility by using the DDR interface will potentially accelerate their adoption, it also introduces performance effects due to the fundamental differences in the underlying DRAM and SCM.
More specifically, the DDR interface specifies that the 64-bit wide channel is driven from a fast SRAM-based \emph{row buffer}, which stores the most recently used row opened from the data array.
Every access to the row buffer is referred to as a \emph{burst}, where the requested word (64 bits) is selected from the row buffer and driven across the interface, which operates at a faster clock rate than the backing arrays.
Accessing an address that is not currently present in the row buffer (i.e., a row buffer miss) means that the existing row must be closed, and the proper row read into the buffer, which is referred to as an \emph{activation}. 
Existing data that is dirty must be written back to the data array prior to opening the new row, in a process that is called \emph{write restoration}.


\input{floats/float-amortization}

Maintaining the same DDR interface and simply swapping the DRAM data array for SCM results in a severe disparity between the channel's speed and the data array's access latency: every row buffer miss now incurs between 4--100$\times$ the latency of DRAM~\cite{suzuki:nonvolatile,wong:stanford} to read the SCM data array.
Given this elevated disparity between the row buffer and data arrays, the bandwidth of modern SCM devices depends heavily on the fraction of accesses that hit in the row buffer.

Figure~\ref{fig:activation_vs_burst} 
graphically demonstrates this behavior, using an example of three write accesses either hitting or missing the same open row buffer. 
We use writes because clean rows do not incur write restorations in persistent memory.
In Figure~\ref{fig:activation_vs_burstA} and Figure~\ref{fig:activation_vs_burstB}, we show the 
increase in total access latency that results from replacing the DRAM array with SCM.
Although the burst time remains the same (due to the standardized DDR interface), SCM's increased activation and write restoration latencies now dominate the overall latency of the three write accesses.
Figure~\ref{fig:activation_vs_burstC} shows how the activation and restoration costs can be amortized when the three writes all hit in the open row, and then are written back together.

Motivated by the performance premium SCM DIMMs place on row buffer hits, we conduct an experiment to compare the average memory access time (AMAT) of a representative DRAM- and SCM-based DDR4-2666 DIMM with 8KB row buffers, varying the size of each memory request. Larger requests serve as a proxy for access patterns that incur more hits in each opened row.
We define AMAT as the average transfer latency of a cache block (64B) from memory to the CPU's last-level cache and model an SCM array with $4\times$ the read latency of DRAM~\cite{micron:ddr4} and $2.5\times$ read/write disparity. Methodology details can be found in \S\ref{sec:method}.

\input{floats/float-scm-bw}

Figure~\ref{fig:AMAT_Comparison} shows the results of this experiment. 
The DRAM's latency quickly becomes bound by the channel's speed, as the 14ns activation time is amortized with approximately 1KB of data transfer.
In contrast, the SCM requires far larger requests to approach the DRAM's AMAT, because of its significantly higher activation time of 60ns.
We therefore conclude that directly replacing DRAM with SCM, using the same DDR interface, places the memory system's performance entirely at the mercy of the applications' access patterns, and whether or not they expose enough row buffer locality.
In the next section, we study typical datacenter applications to determine if their memory access patterns result in the row buffer locality required by SCM DIMMs.

%% file: floats/float-amortization.tex
\begin{figure}[t]
\subfloat{
        \includegraphics[width=\columnwidth]{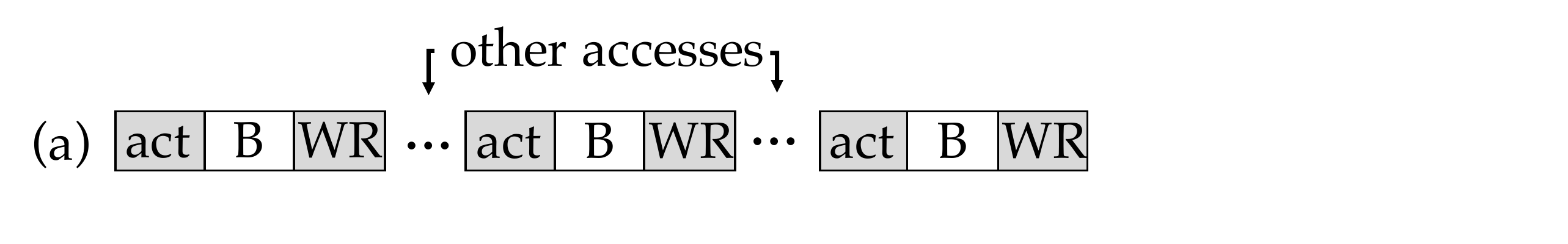}
        \label{fig:activation_vs_burstA}}
\newline
\vspace{-5mm}
\centering
\subfloat{
        \includegraphics[width=\columnwidth]{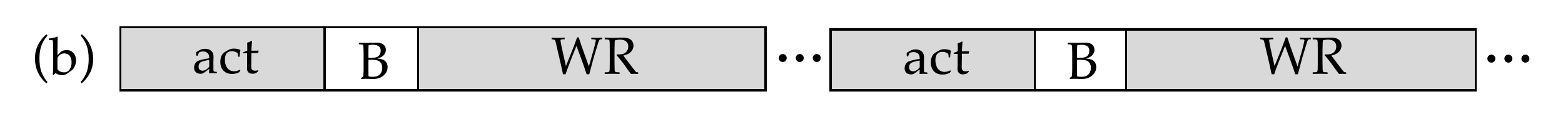}
        \label{fig:activation_vs_burstB}}
\newline
\centering
\subfloat{
        \includegraphics[width=\columnwidth]{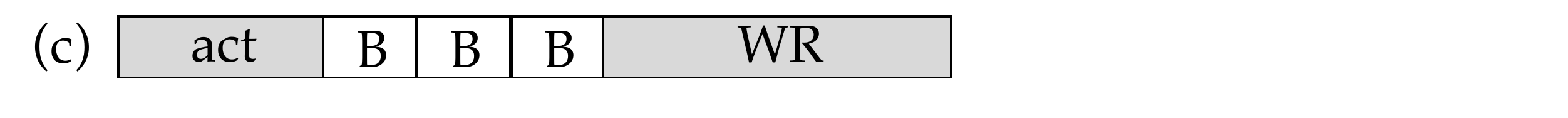}
        \label{fig:activation_vs_burstC}}
\caption{Temporally batched bursts amortize activation cost. \textit{act:} activation, \textit{B:} burst, \textit{WR:} write restoration.\\
(a) DRAM, (b) SCM, (c) SCM with batched bursts.
}
\label{fig:activation_vs_burst}
\end{figure}

%% file: floats/float-scm-bw.tex
\begin{figure}[t]
\centering
\includegraphics[width=\columnwidth]{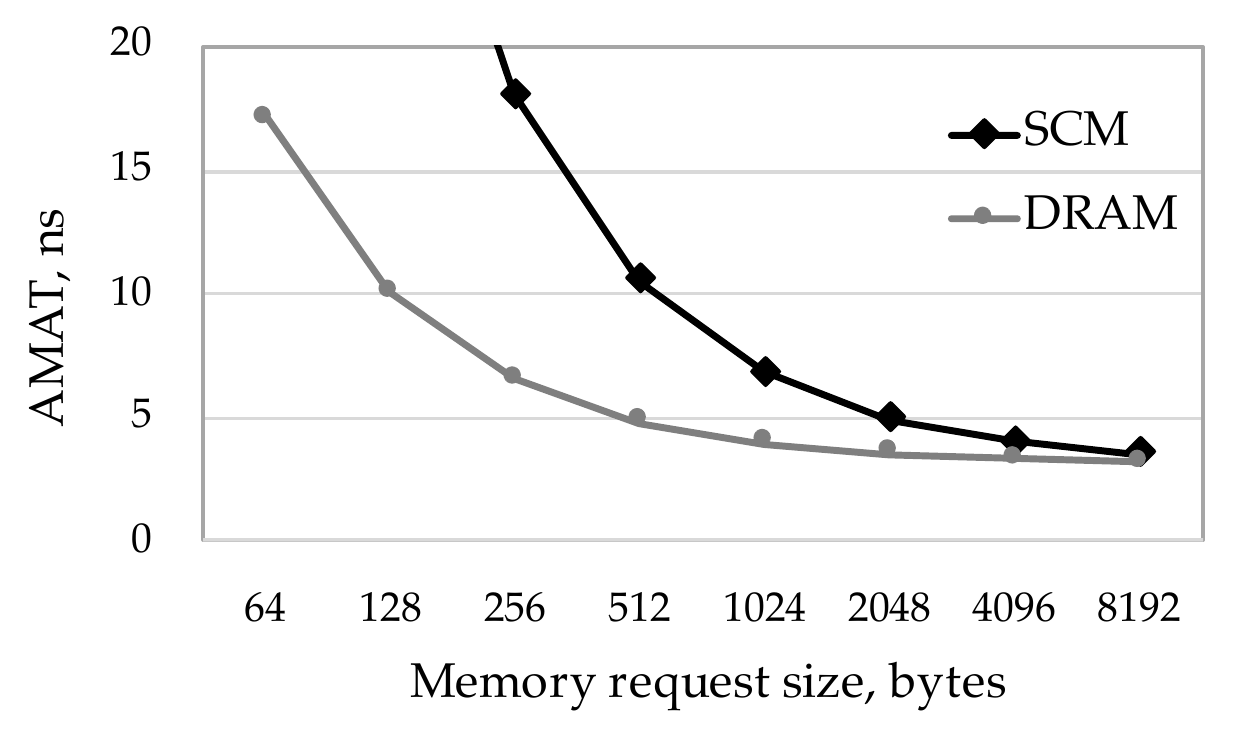}
\caption{AMAT as a function of transfer size (DRAM/SCM row activation time = 14ns/60ns).}
\label{fig:AMAT_Comparison}
\end{figure}

%% file: server_workloads.tex
\section{ SCM Hierarchy Design For Servers}
\newcommand{\DCache}{$3D\$$}
\label{sec:workloads}

In this section, we investigate the feasibility of replacing a server's DRAM with SCM. 
We find that direct replacement results in unacceptable performance degradation, but the use of a modestly sized 3D-stacked DRAM cache makes the replacement viable. 
Due to the high cost and complexity of such 3D stacked caches, we conduct a detailed study to determine the most effective cache organization in terms of capacity, associativity, and block size.

\subsection{Workload compatibility with SCM}
\label{sec:direct_replacement}

\input{floats/float-scm-vs-dram}

As memory latency is a critical performance determinant for datacenter workloads~\cite{ferdman:clearing,kanev:profiling}, the dramatic increase in AMAT caused by replacing DRAM with SCM will directly manifest itself in end-to-end performance degradation.
Therefore, we begin by asking the question: by how much will performance degrade from simply replacing the memory?
We conduct an empirical study of server workloads selected from CloudSuite~\cite{ferdman:clearing}, where we directly compare DRAM-based main memory against an SCM-based alternative.
In this study, we choose to model a latency-optimized SCM device latency-optimized to provide an upper bound on the performance of an SCM DIMM.
Our application performance metric is the number of user-level instructions per cycle (U-IPC) retired by the server during our measurement intervals; U-IPC has been shown to accurately reflect server throughput~\cite{wenisch:simflex}.
For more details on our methodology, see \S\ref{sec:method}.

Figure~\ref{fig:scm_vs_dram} shows the performance of server workloads with SCM-based main memory, normalized to the performance of a server using only DRAM.
The results show that na\"ively replacing DRAM with SCM results in a severe performance degradation of 37\% on average.
To identify whether or not the root cause of this performance degradation is the inflated latency of SCM row activations, as identified in \S\ref{sec:background}, we collect the row buffer hit ratio and the access percentage of each opened row (i.e., number of 64B chunks accessed by the CPU).

We find that 31\% of memory accesses result in a row buffer hit on average, corroborating prior results~\cite{volos:bump}.
Given that we modeled an 8KB row buffer (common in modern DRAM DIMMs~\cite{micron:ddr4}), a 31\% hit ratio corresponds to an average of 2.6KB accessed per row buffer activation (i.e., consumed by the CPU before row closure). 
Our device model in Figure~\ref{fig:AMAT_Comparison} shows that at this request size, an \textit{unloaded} SCM device results in an AMAT 1.33$\times$ higher than DRAM.
A \textit{loaded} system with multiple outstanding requests is expected to have an even higher AMAT, as multiple outstanding requests may be forced to wait in the SCM's command queues, placing even more importance on maximizing row buffer hits.
Indeed, in our experiments running full server workloads (Figure~\ref{fig:scm_vs_dram}), we measure the
AMAT on the SCM-based system to be 2.7$\times$ higher than its DRAM-based counterpart.

In the context of designing a server memory hierarchy using SCM, it is necessary to confirm that the workloads themselves actually have the potential to expose more row buffer locality; if so, a judicious organization of the memory system can immediately restore some of the lost performance.
Prior work that has studied row buffer locality in scale-out workloads has reported that an ideal memory scheduling system can achieve row buffer hit ratios of 77\%~\cite{volos:bump}, a 2.5$\times$ increase over what we observe.
The same work also demonstrated abundant spatial locality present in the workloads themselves, with many pages incurring hits to over 50\% of their constituent cache blocks during their lifetime in the LLC.
Unfortunately, because the memory channels in a modern server processor are multiplexed between many CPU cores, interleaving their access streams destroys a large fraction of whatever row buffer locality would have existed had the application executed in isolation~\cite{volos:bump}.

Since our workloads indeed exhibit row buffer locality that can be harvested to address the concerns discussed in \S\ref{sec:background}, we identify two key requirements for a server wishing to use SCM:
\begin{enumerate}
    \item SCM devices place paramount importance on row buffer hits due to their slow backing data arrays.
    \item Pages that are placed into the DIMM's row buffers need to remain open for long periods of time, in order to collect all of the spatially local accesses, before write restoration occurs.
\end{enumerate}
In the next section, we examine whether any feasible SCM device can meet these requirements.

\subsection{Designing SCMs to meet key requirements}
A na\"ive conclusion from our previous two observations would be that memory architects should build SCM DIMMs with large row buffers to improve the probability of hits, and further optimize memory scheduling to exploit spatial locality therein.
However, the write programming process of various SCM technologies precludes us from constructing rows that are comparable in size to what currently exists in DRAM DIMMs (8KB), due to limitations on write current that can be driven into the data cells during the write restoration stage~\cite{hay:preventing, jiang:fpb}.
Current SCM devices come with row buffer sizes in the 512B--2KB range~\cite{lee:architecting,suzuki:nonvolatile}. 
Even with perfect spatial locality, these smaller rows do not provide enough opportunity to fully amortize the SCM's higher latencies. 
This problem can once again be seen in Figure~\ref{fig:AMAT_Comparison}, where transfer sizes of 1KB and 2KB result in the SCM's AMAT being elevated over DRAM by approximately 1.4$\times$~and~1.3$\times$, respectively. 
Furthermore, techniques for optimizing row buffer locality~\cite{stuecheli:vwq,volos:bump} can only provide a maximum hit rate of ~50\%, which unfortunately lags far behind the hit rate required to provide an equivalent AMAT to DRAM.
These fundamental SCM limitations, combined with limited scheduling scope in the memory controllers, lead us to conclude that SCM cannot serve as a drop-in DRAM replacement.

Although we conclusively show that a server cannot use SCM alone as its main memory while preserving application performance, we reiterate that SCM's desirable characteristics compared to DRAM are capacity and low cost, not raw performance.
Therefore, we propose to learn from existing performance-maximizing solutions that can enable us to recoup some (ideally most) of the performance lost by replacing DRAM with SCM.
Such solutions typically add an additional low latency, high bandwidth memory device, and then seek to serve most memory accesses from the higher performance memory, only relying on the slower memory if necessary.

\input{floats/float-locality-study}

Prior work on DRAM-only systems showed the performance superiority of two-tier memory hierarchies, comprised of a 3D stacked DRAM cache such as Hynix' HBM~\cite{amd:hbm} or Micron's HMC~\cite{micron:hmc}, and a second tier of planar DRAM delivering the necessary memory capacity \cite{volos:fat,jevdjic:die-stacked,qureshi:fundamental}.
In our experiments, a two-tier hierarchy with 3D stacked DRAM as a first tier cache, and conventional planar DRAM as a second tier 
outperforms single-tier planar DRAM by 30\%, corroborating prior work \cite{jevdjic:die-stacked,volos:fat}.

We argue that such a two-tier memory hierarchy is even more applicable to a memory system that includes SCM.
Designed correctly, the first tier can serve a large fraction of memory accesses at DRAM latency and thus provide nearly equal performance to a DRAM only system; building the second tier with SCM provides an order of magnitude more capacity at lower cost than using planar DRAM as suggested in prior work.

In the context of building a high-performance and cost-efficient hybrid DRAM-SCM memory hierarchy, proper design of the first-tier 3D stacked DRAM cache (hereafter abbreviated as {\DCache}) can address the challenges we discussed in \S\ref{sec:direct_replacement}, namely low row buffer hit ratio and increased SCM AMAT.
This is the case for two reasons:
First, a well designed {\DCache} enables the majority of memory accesses to be serviced at DRAM latency rather than requiring an SCM activation.
Second, setting cache block size to be equal to the backing SCM's row buffer size means that the application's spatially localized accesses can be aggregated over the block's relatively long lifetime in the {\DCache}; when the block is evicted and written to the backing SCM, a far greater fraction of the row buffer is actually used than if the row was repeatedly opened and closed in the SCM.
This access coalescing has the same effect as providing near-ideal access scheduling without the requirements for complex reordering logic, and amortizes the SCM's latency.

Having established the critical advantages of using a {\DCache}, we now discuss its design.
We defer a detailed comparison of planar and 3D stacked caches to our evaluation (\S\ref{sec:planar_stacked_comparison}), and a discussion on alternative memory organizations, such as flat DRAM-SCM integration, to \S\ref{sec:related}.

\subsection{3D stacked DRAM cache design}
\label{sec:dram_cache_design}

Due to stacked DRAM's high cost compared to planar DRAM and increased integration complexity, we must be judicious about its architecture and provisioning.
There are three main parameters that define its effectiveness: associativity, capacity, and block size. 
Prior work studying {\DCache}s for server workloads has shown that associativity requirements are modest, with minuscule performance improvements beyond 4 ways \cite{jevdjic:die-stacked}. 
Capacity is a first-order determinant of the cache's filtering efficiency, while block size introduces a tradeoff between leveraging spatial locality and data overfetch. Prior work on {\DCache}s investigated the impact of these parameters for DRAM-based systems \cite{jevdjic:die-stacked,jevdjic:unison,qureshi:fundamental,volos:fat,loh:efficiently}.
We revisit these key design parameters for {\DCache}s in the different context of SCM-based systems, where main memory capacity and access latencies are significantly higher.

To solve the {\DCache} capacity conundrum, we perform an empirical study to investigate whether or not physically feasible {\DCache}s can capture the required working sets of our applications.
We use a trace simulator based on Flexus~\cite{wenisch:simflex}, and conduct a classical miss ratio study where we sweep the {\DCache}'s capacity and search empirically for the "knee of the curve".
We model a fully associative {\DCache} with varied capacity and block sizes, and display the results in Figure~\ref{fig:locality}.

There are two main phenomena that manifest themselves in these results.
First, for all of the block sizes shown, a cache provisioned with approximately 2--4\% of the backing SCM's capacity sits at the knee of the curve and therefore represents the sweet spot for provisioning. 
Recent analysis of Google's production search code~\cite{ayers:memory} corroborates our findings that a similarly sized cache (the authors propose a memory hierarchy with a 1--8GB eDRAM cache, and memory capacity of several hundreds of GBs) can efficiently accommodate the stack, heap and hot data of a multithreaded workload.
Such capacity is reasonable even for die-stacked DRAM technologies, as existing products feature capacities up to 8GB, and industry projections expect 64GB by 2020~\cite{arstechnica:hbm3}.

\input{floats/float-region-density}
\input{floats/float-sensitivity}

Second, we note significantly reduced cache miss ratios with larger {\DCache} block sizes.
For example, Web Search's miss ratio drops from 14.5\% to less than 1\% as the block size increases from 256B to 4KB.
Using larger blocks allows the {\DCache} to amortize the cost of accessing the high-latency backing SCM, as every miss now loads larger chunks of data that will likely be accessed in the future.
We interpret this as further evidence that our set of server workloads exhibits significant spatial locality, but needs a longer temporal window to capture it than the one offered by an open row buffer.
The {\DCache} serves that exact purpose, coalescing accesses within large blocks of data that, upon eviction, amortize the cost of an SCM row activation and write restoration, as illustrated in \S\ref{sec:background}.

Using terminology commonly used in the literature, we argue that DRAM caches should be architected as \emph{page} caches~\cite{jiang:chop,lee:fully,volos:fat} rather than \emph{block} caches, where the term \textit{page} refers to the cache block size being significantly larger than a typical cache block size of 64B. Page-based caches are superior due to the much lower miss ratios exhibited when the cache block size exceeds 1KB.
With a block-based cache, misses to each small block will be serialized once again by the SCM.
Existing {\DCache} designs that use small blocks, typically equal to the L1 cache block size, are unsuitable for SCM-based memory hierarchies~\cite{huang:atcache,loh:efficiently,qureshi:fundamental,sodani:knights}.
Using a page-based {\DCache} solves the problems identified in \S\ref{sec:background}, namely the need to amortize long SCM activations with accesses to spatially local data.

We further justify this choice with a direct study on SCM's latency amortization opportunity as a function of the {\DCache}'s block size.
Figure~\ref{fig:region_density} displays the density of regions being evicted from the {\DCache} and written to the backing SCM, which we define as the fraction of 64B sub-blocks that are accessed during the region's lifetime in the cache.
All of the workloads exhibit similar behavior, albeit grouped into two different clusters.
As the region size increases, density naturally drops.
While most of the workloads exhibit densities exceeding 70\% for region sizes between 512B and 2KB (corresponding to a typical SCM row buffer), Web Serving and Data Analytics have sparser traffic patterns, with ~15\% less density than the others.
Comparing those two workloads to the miss curves in Figure~\ref{fig:locality}, we see that beyond a modest cache size, these same two workloads are the least sensitive to the block size.
For cache sizes large enough to hold $\textgreater$1\% of the dataset, Data Analytics is particularly agnostic to the cache block size, incurring the smallest decrease in miss ratio, due to the fact that it has less innate locality inside each opened row.

By synthesizing the results in Figures~\ref{fig:locality} and~\ref{fig:region_density},
we argue that the {\DCache}~'s block size should match the SCM's row buffer size.
Matching these two parameters allows the {\DCache} to coalesce accesses together and therefore amortize the elevated activation and restoration latencies of the backing SCM. 
Figure~\ref{fig:region_density} essentially shows that the opportunity presented in Figure~\ref{fig:AMAT_Comparison} is attainable, thanks to the combination of the workloads' innate spatial locality with a page-based {\DCache}.

Finally, we present end-to-end application performance results in Figure~\ref{fig:block_sensi} for a system whose memory hierarchy features a {\DCache} sized at 3\% of the backing SCM\footnote{Full methodology details available in \S\ref{sec:method}.}. 
Performance is normalized to a DRAM-based system featuring the same {\DCache} as the SCM-based system.
With the exception of Data Serving, the SCM-based system performs better with larger cache block sizes, until an inflection point appears at 2KB blocks.
This limitation occurs due to overfetching with 4KB blocks, causing bandwidth contention in the SCM, thus setting an upper bound on the {\DCache}'s block size.
Note that the DRAM-based system is less sensitive to the {\DCache}'s block size as a DRAM DIMM's data array latency is much closer to the row buffer access latency as compared to an SCM DIMM. In fact, the DRAM-based system is more sensitive to data overfetch (e.g., Data Serving favors the use of relatively small cache blocks in a DRAM-based system), a problem that is partially offset by the higher benefits of row activation amortization in the case of SCM.

Putting all of our observations together, we present three key design guidelines for memory hierarchies that use SCM-backed {\DCache}s.
First, the performance/cost sweet spot for the {\DCache} is approximately 3\% of the backing SCM's size, and it should necessarily feature large blocks (512B--2KB) to capture the spatial locality present in server workloads and amortize the high SCM access latency. 
We find that for the SCM-backed system, organizing the {\DCache} with 2KB blocks hits the sweet spot between hit ratio and bandwidth misuse because of overfetch, while 1KB blocks result in only marginally lower performance. 

Second, the SCM's row buffer size should be the largest permitted by the underlying memory technology, to maximize the potential of latency amortization, up to a maximum of 2KB to avoid data overfetch (Figure~\ref{fig:block_sensi}).
If the SCM row buffer is smaller than 2KB because of technology limitations, the {\DCache}'s block size should match the SCM row buffer size, as the latter sets the upper bound for SCM latency amortization.

%% file: floats/float-scm-vs-dram.tex
\begin{figure}[t]
        \centering
        \includegraphics[width=\columnwidth]{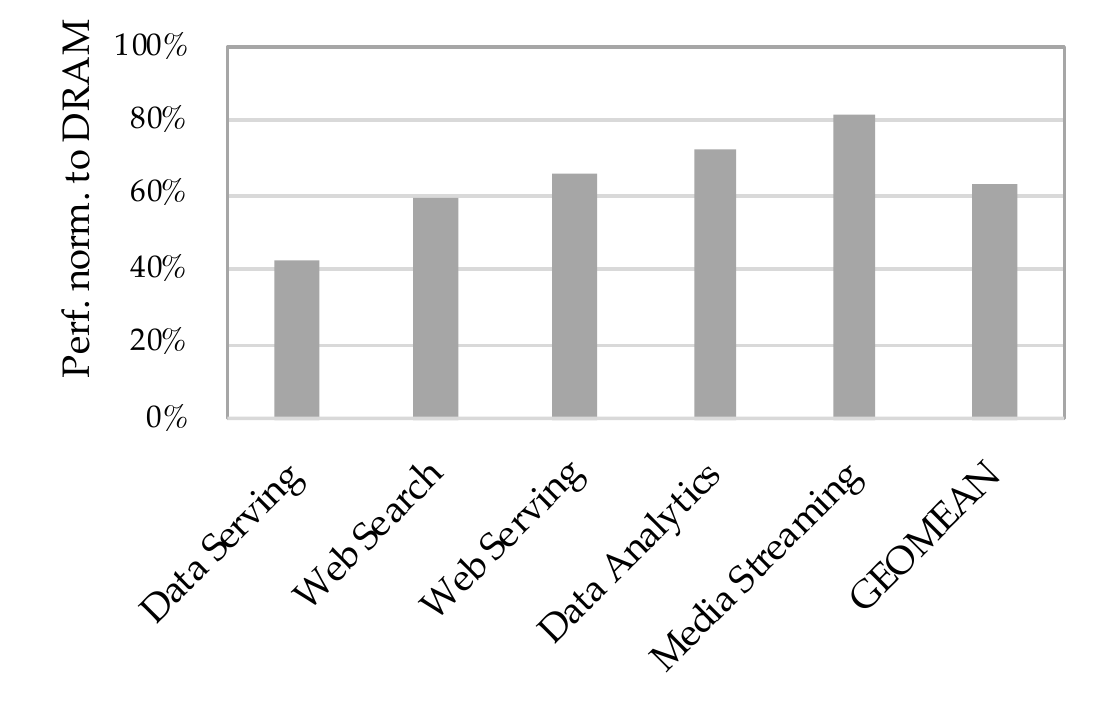}
        \caption{SCM-based vs. DRAM-based memory (SCM Read/Write = 60ns/150ns).}
        \label{fig:scm_vs_dram}
\end{figure}

%% file: floats/float-locality-study.tex
\begin{figure*}[t]
\centering

\subfloat[256B blocks.] {
        \includegraphics[width=0.68\columnwidth]{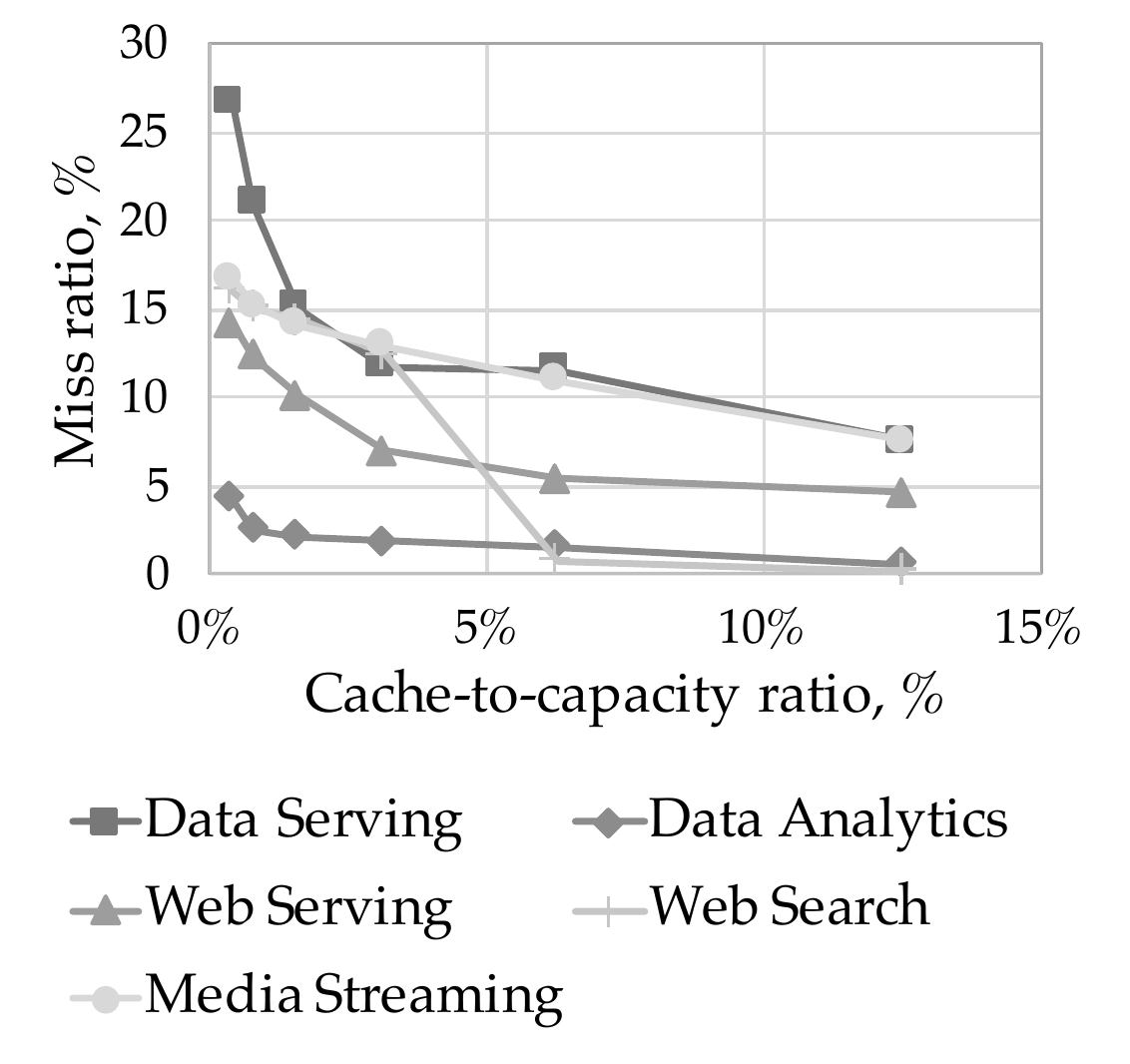}
        \label{fig:256B_locality}}
\centering
\subfloat[1024B blocks.] {
        \includegraphics[width=0.68\columnwidth]{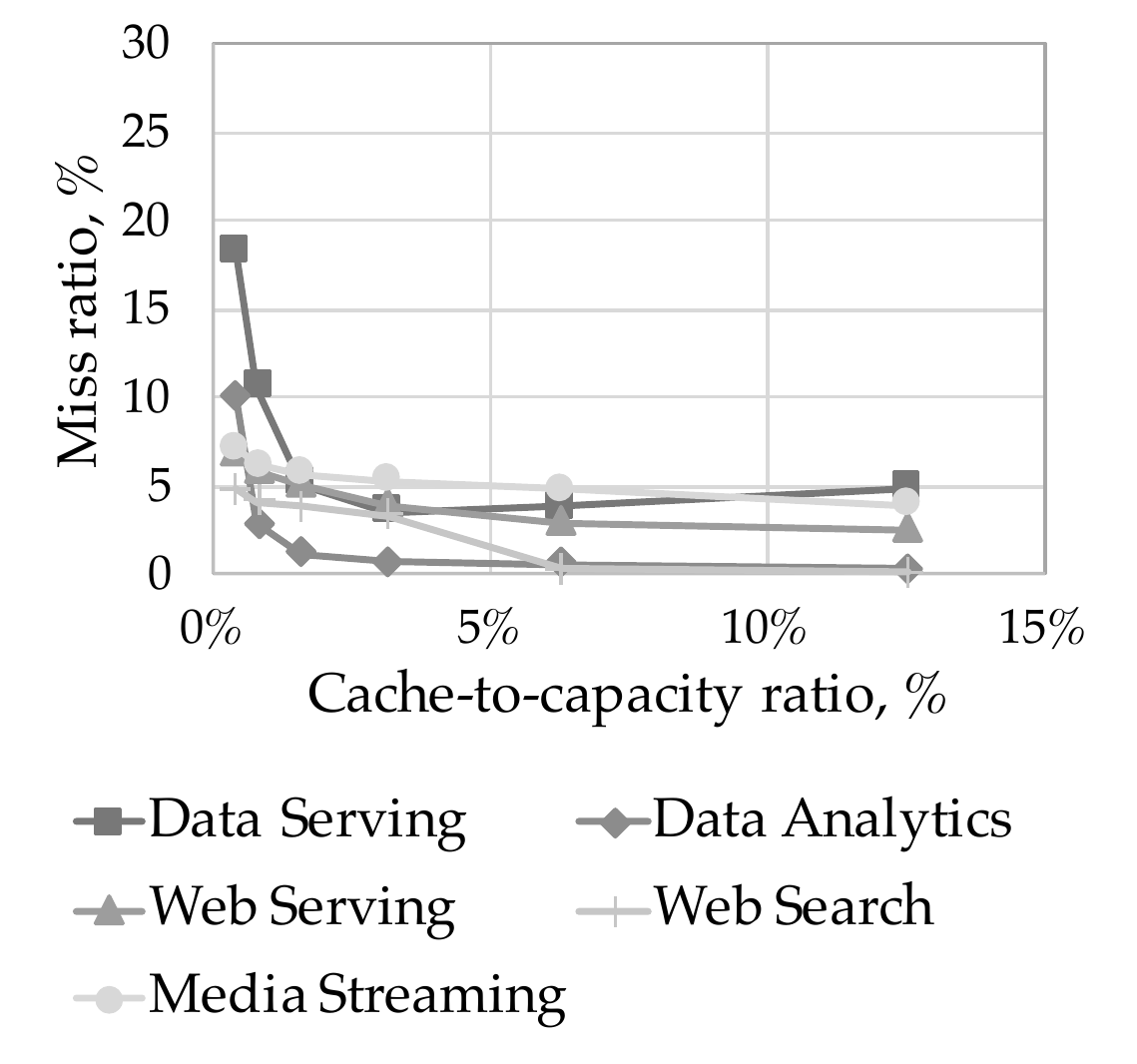}
        \label{fig:1KB_locality}}
\centering
\subfloat[4096B blocks.] {
        \includegraphics[width=0.68\columnwidth]{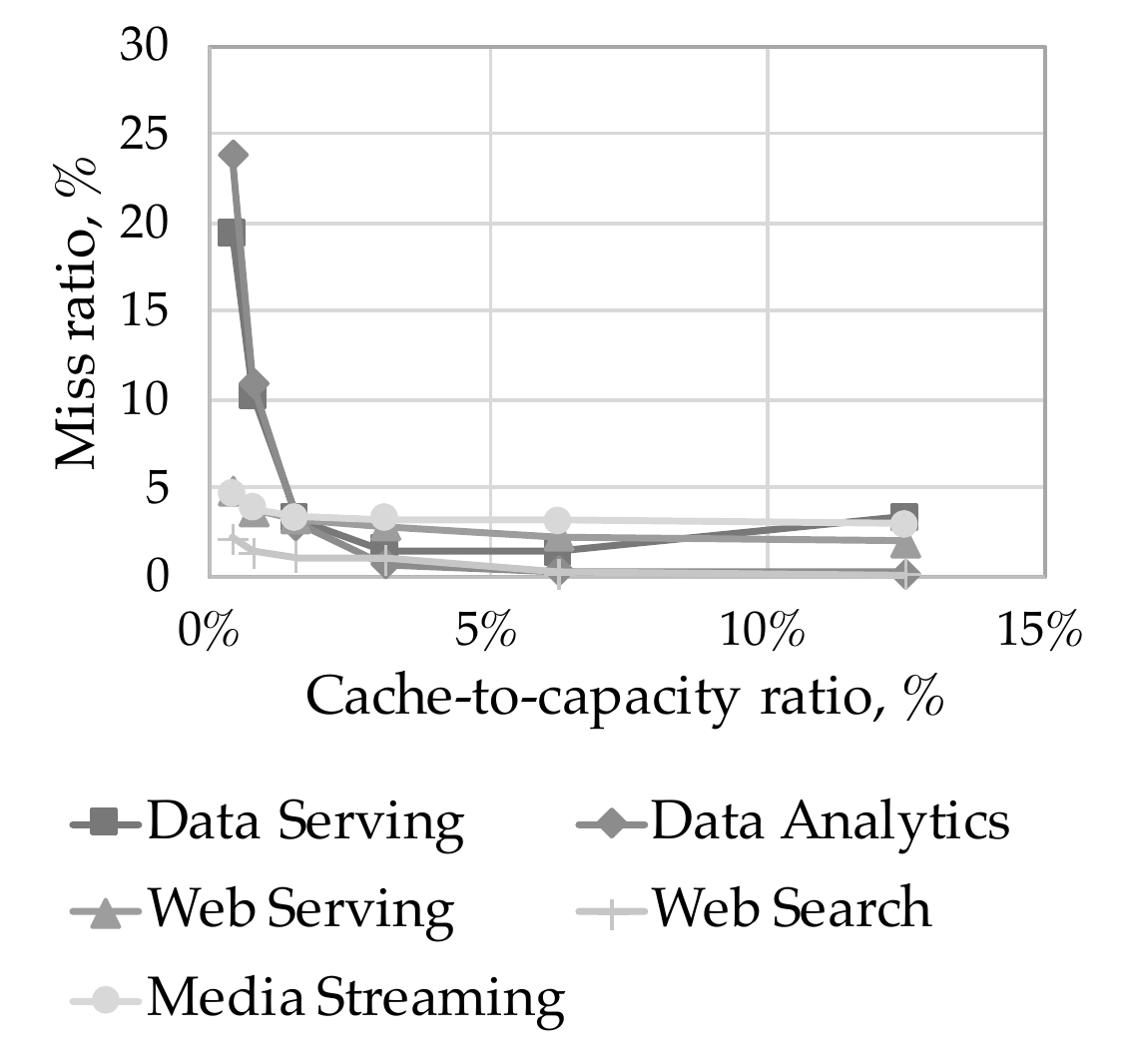}
        \label{fig:4KB_locality}}

\caption{Miss ratios for a fully associative DRAM cache. The x-axis sweeps through the capacity ratio between the DRAM cache and the backing memory.}
\label{fig:locality}
\end{figure*}

%% file: floats/float-region-density.tex
\begin{figure}[t]
\centering

        \includegraphics[width=\columnwidth]{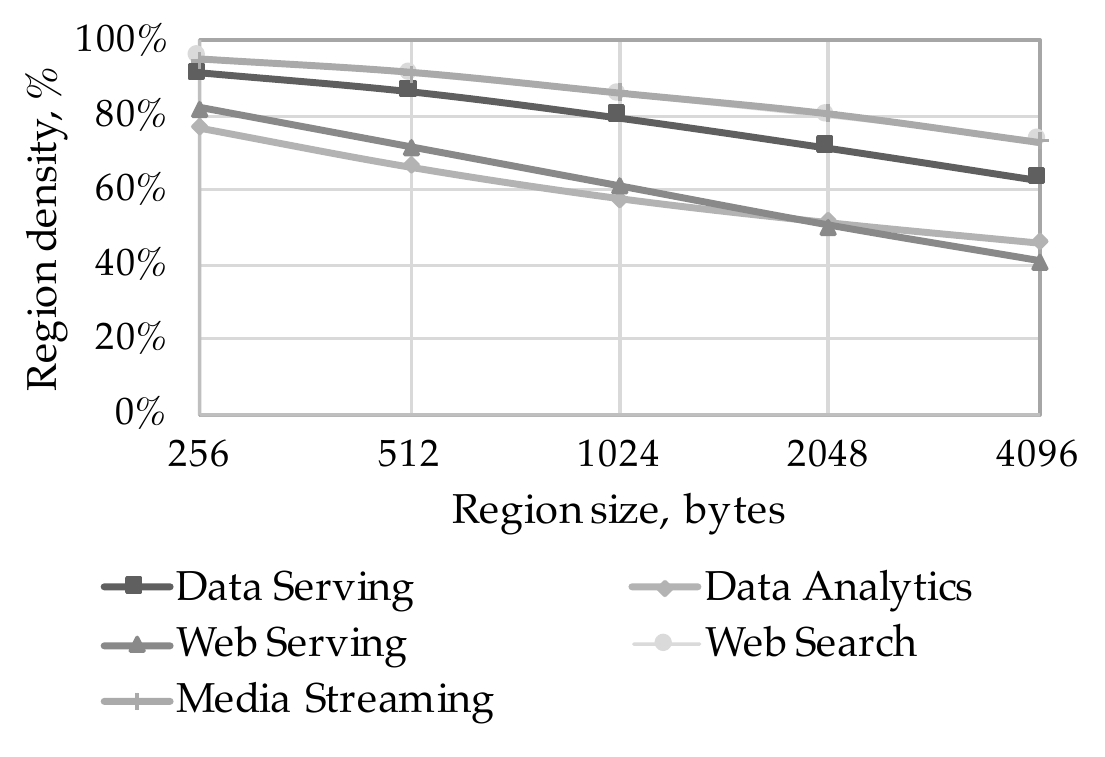}
        \label{fig:density}
\caption{Percentage of 64B sub-blocks in each DRAM cache block (region size) during its lifetime, measured at first sub-block's eviction.
Note: Web Search and Media Streaming lines overlap.
}
\label{fig:region_density}
\end{figure}

%% file: floats/float-sensitivity.tex
\begin{figure*}[t]
\centering
        \includegraphics[width=2.1\columnwidth]{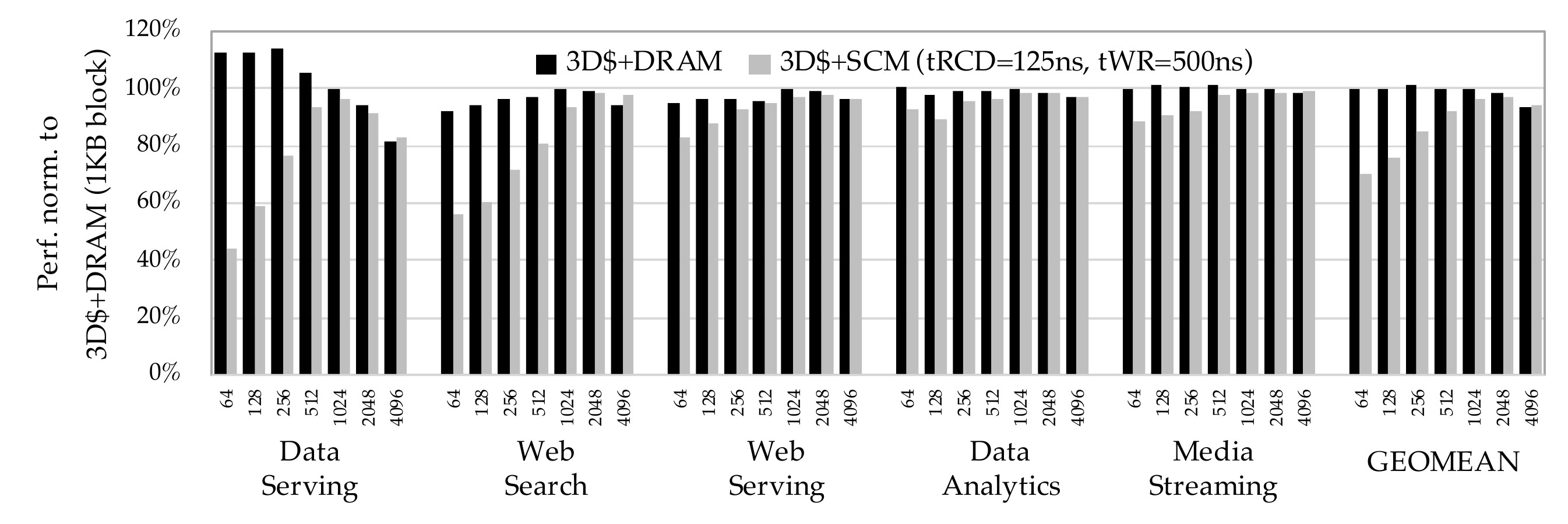}
\caption{
Performance of DRAM-based (black) and SCM-based (gray) systems with the same {\DCache} of variable block size.} 

\label{fig:block_sensi}
\end{figure*}

%% file: scm_space.tex
\section{SCM Cost/Performance Tradeoff Exploration}
\label{sec:approaches}

\input{floats/float-model}


In the previous section, we demonstrated that an SCM-based system is able to attain competitive performance with a DRAM-based one, thanks to the addition of a {\DCache}. 
However, 3D stacked DRAM technology costs at least an order of magnitude more per bit than SCM, conflicting with the initial motivation of replacing DRAM with SCM as a more cost-efficient memory.
Hence, whether the resulting memory hierarchy represents an attractive solution depends on whether the cost reduction from replacing DRAM with SCM offsets the additional cost of a DRAM cache. 
As SCM itself is a technology that offers a broad spectrum of density/cost/performance operating points, the challenge is to minimize its cost while preserving its performance at acceptable levels. 
In this section, we trim the broad design space by identifying the key parameters that define SCM's performance in the context of our memory hierarchy.


In general, the denser the SCM, the lower its cost per bit \cite{lu:tutorial,qureshi:morphable,stanisavljevic:demonstration}. We therefore use SCM density as a proxy for SCM cost. 
The goal is to deploy the densest (and therefore cheapest) possible SCM while respecting end-to-end performance goals.
Unfortunately, common density optimizations like storing multiple bits per cell or vertical stacking of multiple cell layers result in higher access latency, lower internal bandwidth, and potentially higher read/write disparity~\cite{lu:tutorial,wd:reram,suzuki:nonvolatile,wong:stanford,xu:understanding}.
Therefore, solving the cost-performance optimization puzzle requires SCM designers to understand which parameters affect end-to-end performance the most, and by how much.

We identify read latency, write latency, and row buffer size as the three SCM design parameters that control end-to-end application performance. 
Read latency (i.e., SCM row activation delay) sits on each memory access' critical path. 
Write latency (i.e., write restoration delay), even though off the critical path, may cause head-of-line blocking delays inside the SCM DIMM \cite{arjomand:boosting,qureshi:improving}.
Finally, as discussed previously, the row buffer size defines the extent to which SCM's high access latency can be amortized.

Putting all three parameters together---row buffer size, read latency, and write latency---we devise a three-dimensional SCM design space, illustrated in Figure~\ref{fig:model_3d}. 
All the SCM configurations that satisfy the performance target reside inside the volume shaped like a triangular frustum. 
The SCM devices with the lowest read and write latencies lie close to the vertical axis. 
All designs for a given row buffer size are represented by a horizontal cut through the frustum, and the resulting plane indicates the space of all read and write latencies that are tolerable with that row buffer size. 
The frustum's lower base is defined by the smallest row buffer size that is sufficient to amortize the SCM's row activations (\S\ref{sec:workloads}); on that plane, only the fastest SCM devices are acceptable, which are unlikely to deliver the desired high density and low cost.
Growing the row buffer size (and implicitly the SCM's internal bandwidth) widens the design space, as increased amortization opportunity reduces the overall system's sensitivity to high activation latency.

Given a target workload's characteristics, our methodology helps device designers reason about the feasibility of employing different SCM technologies as main memory. 
For example, with multi-level cells, designers may deploy smaller serial sensors to optimize for higher density, by sacrificing read latency \cite{xu:understanding}.
Another example is the write latency/bandwidth tradeoff, where designers may choose a different cell writing algorithm, optimizing either for low latency or high bandwidth based on their performance needs. 
Fewer high-current write iterations result in faster writes, but place an upper limit on the row buffer size because of fundamental limitations on the current that can be driven through the data array at any given time~\cite{zhang:balancing}. 
A general observation from our design space exploration is that devices with bigger row buffers are appealing, as they widen the design space and offer better opportunities for latency amortization. 
However, benefiting from this characteristic requires the target application domain's workloads to exhibit a certain degree of spatial locality.

To summarize, we analyzed three SCM parameters, namely the row buffer size and read/write latencies, and devised a methodology that prunes the vast SCM design space and finds best-fitting solutions for a given application domain.
In the following sections, we instantiate this model for our set of server workloads and evolve it from qualitative to quantitative, selecting parameters representative of existing SCM technologies. 
We then use our model to perform a case study of four representative PCM configurations, and compare the performance/cost metric of memory hierarchies built from each configuration.

%% file: floats/float-model.tex
\begin{figure}[t]
\centering

        \includegraphics[width=.95\columnwidth]{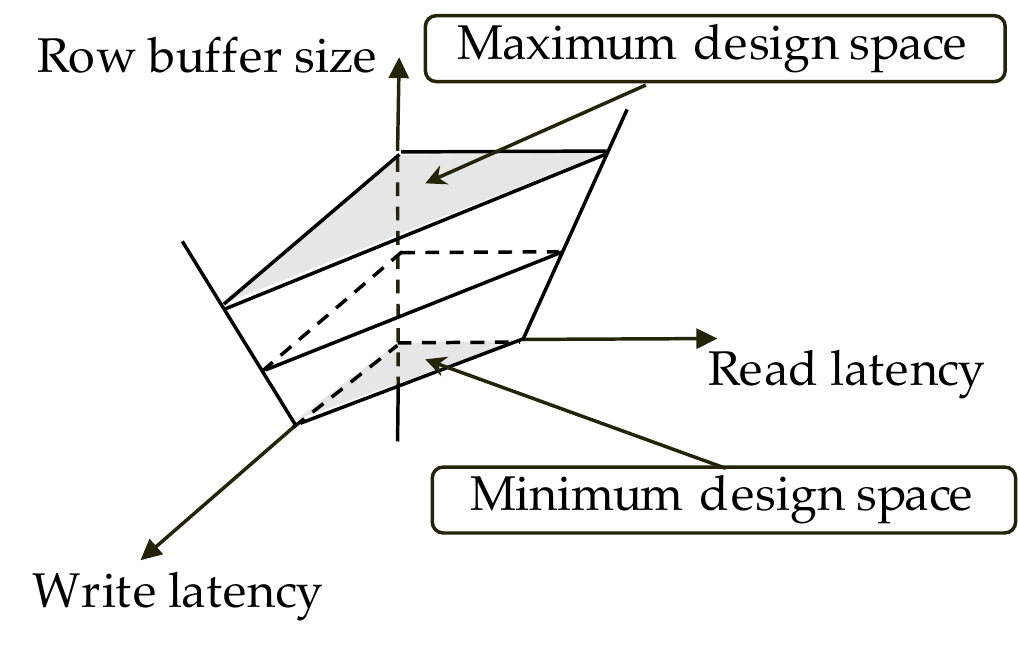}
\caption{Performance model for SCM design space exploration. 
}
\label{fig:model_3d}
\end{figure}

%% file: methodology.tex
\section{Evaluation methodology}
\label{sec:method}

In this section, we describe the organization of each system we model throughout the paper, provide the details of our simulation infrastructure, state our performance and cost assumptions, and finally list the parameters we use for our case study with PCM.\\

\noindent\textbf{System organization.} Next-generation server processors will feature many cores to exploit the abundant request level parallelism present in online services \cite{ferdman:clearing, lotfi-kamran:scale-out}. 
Recent server chips follow this very trend: AMD's Epyc features 32 cores per socket~\cite{mpr:epyc}, Qualcomm's Centriq 48 cores~\cite{qualcomm:centriq}, and Phytium's Mars 64 cores~\cite{zhang:mars}. 
To make simulation turnaround time tractable, we model a server with 16 cores and a single memory channel, representing a scaled-down version of the OpenCompute server roadmap, which calls for 96 cores and 8 memory channels~\cite{opencompute}.

We configure the DRAM cache's size as 3\% of the workload's dataset in order to achieve the cache-to-memory-capacity ratio required for satisfactory performance (see \S\ref{sec:dram_cache_design}), unless specified otherwise. 
We model a 4-way set-associative cache, and for each evaluated configuration, we set the DRAM cache's block size equal to the SCM's row buffer size.
The DRAM cache is connected to the chip over a high-bandwidth interface (e.g., a SerDes serial link or HBM-like silicon interposer)~\cite{jedec:wideio2,micron:hmc,amd:hbm}, which in turn is connected to the main memory over a conventional DDR4-2666 channel~\cite{micron:ddr4}.
A block diagram of our modeled system is displayed in Figure~\ref{fig:modeled_system}.\\


\input{floats/float-modeled-system}

\noindent\textbf{Workloads.} Our server workloads are taken from CloudSuite \cite{ferdman:clearing}: Data Serving, Web Search, Media Streaming, Data Analytics, and Web Serving.
We measure performance by collecting the server's User-level IPC (U-IPC), which is defined as the ratio of user instructions committed to the total number of cycles spent in both user and kernel spaces. 
Prior work~\cite{wunderlich:smarts} has shown U-IPC to be a metric representative of application throughput.
We use the rigorous SMARTS sampling methodology~\cite{wunderlich:smarts} to compute all of our performance values using a 95\% confidence interval of 5\% error.


For each workload, we configure the overall memory capacity (i.e., second tier of the hierarchy) to be equal to the workload's dataset size (i.e., Data Serving, Web Search and Media Streaming have 16GB datasets, while Data Analytics and Web Serving have 32GB datasets).
However, today's datacenter-scale applications can have much larger datasets that even span into the terabyte range~\cite{ayers:memory,novakovic:case}; since our work makes specific claims about the capacity ratio relating the two tiers of our memory hierarchy, we conducted a study to verify that our results stand for larger datasets.

To confirm the validity of our results as the dataset size scales up, we analytically studied the relationship between hot and cold data as the entire dataset size increases by orders of magnitude.
A key input to these models is a representative query distribution that accurately reflects the skewed popularity phenomenon in datacenter applications.
We used the canonical Zipfian distribution, commonly used to rank the frequency of distinct items in a massive dataset~\cite{novakovic:case,cha:i,armstrong:linkbench,sharma:blink,fan:small,venugopalan:beehive}. In this experiment, we arbitrarily define the \textit{hot fraction} of the dataset as the subset of items that absorbs 70\% of the accesses.

We studied Zipf coefficients ($\alpha$) from 0.6 to 1.01, and observed that while the absolute dataset size scales, the fraction of the dataset classified as hot decreases. 
This means that our choice to size the {\DCache} as a fraction of the total dataset is actually a conservative choice; larger datasets will have smaller hot fractions, which will be absorbed by the {\DCache}.
For example, given $\alpha$=0.9, scaling a 50 million object dataset by 100-fold leads to a slight decrease of the hot fraction from 5.5\% to 4.3\% in our analytical model.
Therefore, we expect that our scaled down system's performance is representative of applications with larger datasets, as increasing the absolute size does not significantly affect the disparity between hot and cold data, and actually leads to even higher data locality.
This phenomenon would result in a {\DCache} that is an even smaller fraction of the backing memory's capacity than what we assumed so far.\\

\noindent\textbf{Simulation infrastructure.} We use the Flexus \cite{wenisch:simflex} full-system cycle-accurate simulator coupled with DRAMSim2~\cite{rosenfeld:dramsim2}. 
To extend DRAMSim2 to support non-uniform SCM access latencies, we adjusted its $t_{RCD}$ and $t_{WR}$, and added SCM-related parameters ($t_{RRDpre}$ and $t_{RRDact}$, similarly to the models used by prior work \cite{arjomand:boosting,lee:architecting}). 
To simplify our explanations, we refer to the \textit{read} and \textit{write} latencies of the SCM device as $t_{RCD}$ and $t_{WR}$, as they define the major part of the data array's access. 

Without loss of generality, we consider a DRAM cache with its tags stored in SRAM, a common design choice in prior work~\cite{volos:fat,jevdjic:die-stacked,jiang:chop,huang:atcache,chou:candy}.
For the DRAM cache's memory controller, we use a critical-block-first policy and FR-FCFS open-row scheduling with page-based interleaving. 
We assume that each SCM is packaged in a DIMM form factor. 
To model different SCM configurations, we replicate expected performance and cost characteristics from recent prototypes~\cite{jurczak:advances,wd:reram,wong:stanford,computerworld:3dxpoint,zhou:hnvm,lu:tutorial,suzuki:nonvolatile}.
For the SCM's controllers, we model an open-row policy, FR-FCFS scheduling, and page-based interleaving, which is optimized for bulk transfers (\S\ref{sec:background}).
The write buffer's size corresponds to the number of banks, with each write entry equal to the page size. 
Finally, even though existing SCM devices feature row buffers up to 2KB~\cite{suzuki:nonvolatile}, we extend our study to 4KB, which is the largest region we expect to capture significant spatial locality (assuming a 4KB OS page size).
Table~\ref{table:flexusSetup} summarizes our simulation parameters.


\input{floats/float-flexus-tbl}

\subsection{Phase-change memory assumptions} 
\label{sec:pcm_assumptions}

PCM is generally considered the most mature SCM technology, as its performance, density and endurance characteristics are well-studied.
Additionally, industry has built reliable single-level and multi-level cell (up to 3 bits/cell)  configurations. We assume a typical PCM cell and project its performance characteristics for single-level (SLC), multi-level (MLC) and triple-level cells (TLC), which store 1, 2, and 3 bits/cell respectively.
For the baseline SLC-PCM configuration, we assume 60ns read latency, and 150ns write latency. Based on a survey of recent PCM prototypes~\cite{suzuki:nonvolatile}, we assume a maximum row buffer size in SLC-PCM of 1024B.



Assuming the same cell material, we project MLC-PCM to operate with 120ns read latency, and a range of possible write latencies, depending on the algorithm used for cell writing.
Prior work~\cite{zhang:balancing} has described two ways to program an MLC cell. The first approach, which we call $MLC_{lat}$, favors faster writes, resulting in write latencies of 550ns and $512$B row buffers. The second approach, which we call $MLC_{BW}$, favors higher bandwidth, resulting in write latencies of $1000$ns and $1024$B row buffers. Finally, we project the specifications of TLC-PCM based on a recent industrial prototype~\cite{athmanathan:multilevel,stanisavljevic:demonstration}, and assume read and write latencies of $250$ns and $2350$ns, respectively. For the row buffer size, we optimistically assume $512$B.\\




\input{floats/float-pcm-tbl}

\noindent\textbf{Cost model.} To evaluate the cost of the memory subsystem, we build a model for both planar and 3D stacked DRAM, as well as SCM of different densities. 
We compare different technologies according to their expected cost/bit metric, normalizing to the same total capacity.
Taking planar DRAM's cost/bit as a baseline, we project 3D stacked DRAM's cost/bit to be 7$\times$ higher than planar DRAM, as cooling and bonding costs increase for stacked dies~\cite{dong:fabrication}. We discuss the implications of possible 3D stacked DRAM cost changes over time in \S\ref{sec:disc}.

Due to the higher manufacturing costs because of the immaturity of PCM technologies~\cite{computerworld:3dxpoint,zhou:hnvm}, we conservatively assume that the cost/bit of SLC-PCM is equal to commodity planar DRAM. Then, we assume cost reductions for MLC and TLC-PCM proportional to the number of stored bits per cell (i.e., 50\% and 25\% of the cost for 2 and 3 bits/cell). Table~\ref{table:pcm} summarizes the performance and cost assumptions for all considered technologies: planar DRAM, stacked DRAM, and the four aforementioned PCM configurations (SLC, $MLC_{lat}$, $MLC_{BW}$, and TLC).

%% file: floats/float-modeled-system.tex
\begin{figure}[t]
        \centering
        \includegraphics[width=0.9\columnwidth]{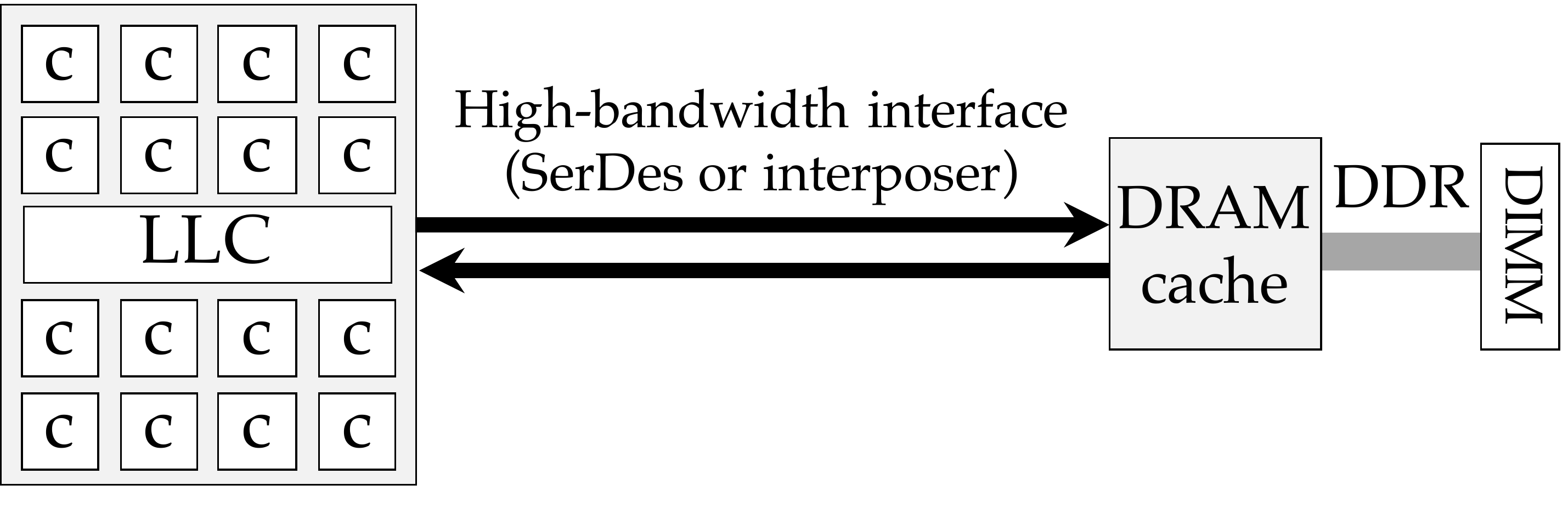}
        \caption{Modeled system overview.}
        \label{fig:modeled_system}
\end{figure}

%% file: floats/float-flexus-tbl.tex
\begin{table}
\begin{center}
\begin{footnotesize}
\def\arraystretch{1.2}	
\begin{tabular}{ |>{\centering\arraybackslash} p{30mm} | p{47mm} |}
\hline
\multirow{3}{*}{Cores}  &   ARM Cortex-A72-like; 64-bit, 2.5GHz, \\
        & OoO, 128-entry ROB, TSO, \\  
		& 3-wide dispatch/retirement\\ 
\hline
\multirow{3}{*}{L1 Caches} & 32KB 2-way L1d, 48KB 3-way L1i, \\
               &  64-byte blocks, 2 ports, 32 MSHRs, \\
	  		   & 3-cycle latency (tag+data) \\
\hline
\multirow{2}{*}{LLC} & Shared block-interleaved NUCA, 16-way,\\
                     & 4MB total, 1 bank/tile, 8-cycle latency \\
\hline
Coherence & Directory-based Non-Inclusive MESI\\
\hline
{Interconnect} & 16$\times$8 crossbar, 16B links, 5 cycles/hop \\
\hline
{DRAM cache}  & 4-way, SRAM-based tags, 20ns lookup \\
\hdashline
{Planar}      & DDR4-2666, 8192B row buffer \\
\hdashline
{3D stacked ({\DCache})}  & SerDes @10GHz, 160Gb/s per direction \\
\hdashline
$t_{CAS}$-$t_{RCD}$-$t_{RP}$-$t_{RAS}$-$t_{RC}$ & 14-14-14-24-38\\
$t_{WR}$-$t_{WTR}$-$t_{RTP}$-$t_{RRD}$ & 9-6-3-3\\
\hline
\multirow{4}{*}{Main memory} & 32GB, single memory channel, \\
& 2 ranks, 8 $\times8$ banks per rank, \\
& Memory controller: 64-entry queue,  \\
& chan:row:bank:rank:col interleaving\\
\hline
Planar DRAM & DDR4-2666, 8192B row buffer \\
\hdashline
$t_{CAS}$-$t_{RCD}$-$t_{RP}$-$t_{RAS}$-$t_{RC}$ & 14-14-14-24-38\\
$t_{WR}$-$t_{WTR}$-$t_{RTP}$-$t_{RRD}$ & 9-6-3-3\\
\hline
SCM & DDR4-2666, 512--4096B row buffer \\
\hdashline
$t_{CAS}$-$t_{RCD}$-$t_{RP}$-$t_{RAS}$-$t_{RC}$ & 14-$t_{read}$-14-24-$t_{read}$\\
$t_{WR}$-$t_{WTR}$-$t_{RTP}$ & $t_{write}$-6-3\\
$t_{RRDpre}$-$t_{RRDact}$ & 2-11 \\
\hline
\end{tabular}
\caption{System parameters for simulation on Flexus. 
Timing parameters for all memory technologies shown in ns.}
\label{table:flexusSetup}
\end{footnotesize}
\end{center}
\end{table}

%% file: floats/float-pcm-tbl.tex
\newcolumntype{P}[1]{>{\centering\arraybackslash}p{#1}}

\begin{table}
\begin{center}
\begin{footnotesize}
\def\arraystretch{1.2}	
\begin{tabular}{ |>{\centering\arraybackslash} p{16mm} | P{7mm} | P{7mm} | P{14mm} | P{12mm} | P{6mm} |}
\hline
Cell configuration & Read lat., ns & Write lat., ns & Total banks/ device & Row buffer, bytes & Cost/ bit, \% \\
\hline
Planar DRAM     & 14    & 9     & 16  & 8192  & 100   \\
\hline
Stacked DRAM    & 14    & 9     & 512 & 256   & 700   \\
\hline
SLC             & 60    & 150   & 16  & 1024  & 100   \\
\hline
$MLC_{lat}$     & 120   & 550   & 16  & 512   & 50    \\
\hline
$MLC_{BW}$      & 120   & 1000  & 16  & 1024  & 50    \\
\hline
TLC             & 250   & 2350  & 16  & 512   & 25    \\

\hline

\end{tabular}
\caption{PCM performance and cost characteristics.}
\label{table:pcm}
\end{footnotesize}
\end{center}
\end{table}

%% file: evaluation.tex
\section{Evaluation}
\label{sec:eval}

Our methodology seeks to quantify the SCM design space model that we developed in \S\ref{sec:approaches}, and validate the performance of our proposed memory hierarchy.
Using our simulation infrastructure, we study a variety of combinations of row buffer sizes and read/write latencies that we gathered from device datasheets, industry projections, and published literature \cite{suzuki:nonvolatile,athmanathan:multilevel,stanisavljevic:demonstration,lu:tutorial}.
We then conduct a case study investigating the feasibility of four different PCM configurations from both performance and cost perspectives, based on the assumptions summarized in Table \ref{table:pcm}.
Finally, we demonstrate that a {\DCache} not only results in better performance but also improved performance/cost as compared to a planar DRAM cache, when used as the first tier of an SCM hierarchy.

\subsection{Quantifying SCM design space for servers}

Figure~\ref{fig:model_2d} superimposes a number of different horizontal cuts of Figure~\ref{fig:model_3d}'s triangular frustum. 
Each point represents an SCM configuration with different read and write latencies. Each different row buffer size configuration is depicted by a diagonal line that separates the configurations that satisfy the performance target from those that do not (i.e., design points that fall inside or outside the frustum's volume).
Similarly to prior work evaluating emerging technologies in datacenters \cite{agarwal:thermostat,gao:network},
we set the bound of acceptable performance for the SCM-based memory hierarchy to be within 10\% of the best DRAM-based system---which features a {\DCache}~with 1KB blocks---for every one of the evaluated workloads. 


In Figure~\ref{fig:model_2d}, the points below each diagonal line satisfy the performance target. 
For example, with a row buffer size of 512B, the slowest configurations that match the performance target are the skewed SCM configuration with 125ns read and 500ns write latencies, and the symmetric configuration of 250ns read and write latencies. 

\input{floats/float-model-eval}
\input{floats/float-model-pcm}

As we explained in \S\ref{sec:approaches}, the row buffer size sets the upper bound for SCM access latency amortization, and is therefore  the parameter implicitly defining the highest SCM latencies that can be tolerated. 
Increasing the row buffer size from 512B to 2KB expands the design space linearly. Hence, the maximum read and write latencies meeting our performance target increase proportionally.
For example, sweeping the row buffer size from 1KB to 2KB, the maximum acceptable read latency increases from 250ns to 500ns, while the maximum allowed write latency grows from 1 to 2\si{\micro}s. 
This relation between the maximum allowed latency and  row buffer size demonstrates the efficiency of amortizing longer SCM latencies over multiple accesses within a large row buffer.

Application performance turns out to be much less sensitive to slow writes, as compared to reads, 
because writeback traffic is not directly on the critical path of memory access. 
This leads us to the important conclusion that SCM's inherent read/write performance disparity is a secondary concern for hierarchical designs, as a carefully organized {\DCache} collects writes to pages and then drains them to SCM in bulk upon eviction.
Accessing the SCM in bulk allows the system to tolerate these elevated write latencies.

Growing the row buffer and {\DCache}'s block size beyond 2KB is not worthwhile, since some applications do not take advantage of the additional data fetched.
For example, Data Serving fails to satisfy our performance target using blocks larger than 2KB, even for DRAM-based systems, as we have seen before in Figure~\ref{fig:block_sensi}. For the rest of the workloads, growing the row buffer size to 4KB widens the design space further, up to 1\si{\micro}s read and 4\si{\micro}s write latencies (not shown on Figure \ref{fig:model_2d}). 
However, most of the workloads experience performance degradation with {\DCache} blocks and SCM row buffers of 4KB, as compared to the corresponding 2KB configuration. For example, for the skewed configuration with 125/2000ns read and write latencies, increasing the row buffer from 2KB to 4KB leads to mean performance degradation of 3\% and up to 9\% for Data Serving. As a result, designers may consider using slower memory with a row buffer size bigger than 2KB only if their applications exhibit that amount of spatial locality.

To summarize, we quantified the frontier that separates plausible SCM configurations from those that are not able to reach the performance target. 
We demonstrated that a bigger row buffer and corresponding {\DCache} block size widen the SCM design space, albeit without exceeding the spatial locality exhibited by the applications' access patterns (2KB for our set of server applications). 
Finally, we make the observation that a simple page-based design efficiently mitigates conventional SCM read/write latency disparity, eliminating the need for any additional disparity-aware mechanisms. 

\subsection{Case study with phase-change memory}
\label{sec:pcm_case_study}

We now demonstrate the utility of our performance model by using it to reason about the implications of a number of plausible PCM configurations on overall system performance and cost. We evaluate the economic feasibility of the SLC, $MLC_{lat}$, $MLC_{BW}$ and TLC PCM configurations we introduced in \S\ref{sec:pcm_assumptions}. 

Figure~\ref{fig:pcm_model2d} shows all four configurations as points, according to their assumed read and write latencies. 
Points with no fill represent configurations with a 512B row buffer, while filled points depict configurations with a 1024B row buffer. 
Similarly to Figure~\ref{fig:model_2d}, diagonal lines bound the configurations that match our performance target (within 10\% of the best DRAM-based system for each workload), according to their corresponding row buffer sizes. 
For all the configurations, we model an SCM hierarchy with a page-based {\DCache}, sized at 3\% of the application dataset, and organized in pages equal to the row buffer size. 
However, as the TLC-PCM based hierarchy fails to deliver acceptable performance with a {\DCache} sized at 3\% of the PCM, we also evaluate TLC-PCM with 6\% and 12\% {\DCache}s, as the low price of TLC-PCM (25\% of DRAM) allows us to consider larger {\DCache}s.
Table~\ref{table:pcm_res} summarizes the performance results and overall memory hierarchy cost for each PCM technology we considered, normalized to a planar DRAM configuration.

The SLC-PCM configuration we consider attains performance within 2\% of the best DRAM configuration with a {\DCache}. 
Although well within our performance target, SLC-PCM's cost/bit is too high to offset the expense of adding the {\DCache}.

\input{floats/float-pcm-res-tbl}

For MLC-PCM, we consider two alternatives: $MLC_{lat}$ and $MLC_{BW}$, which are optimized for low write latency and high internal bandwidth respectively. The row buffer sizes of these configurations are 512B and 1KB. According to the model in Figure~\ref{fig:pcm_model2d}, both configurations deliver performance within the 10\% performance target. Although $MLC_{lat}$ outperforms $MLC_{BW}$ by 3\% on average (1.28 vs. 1.24), designers may prefer $MLC_{BW}$ as its lifetime is a few orders of magnitude longer~\cite{zhang:balancing}.
As the cost/bit of MLC-PCMs is half that of planar DRAM, the overall cost and performance/cost metrics improve by 40\% and 66\% as compared to the DRAM-based system with a {\DCache} of the same capacity  (1.78/1.72 vs. 1.07). As compared to planar DRAM, MLC-PCM improves performance/cost by 1.7--1.8$\times$, reducing overall memory cost by 28\%.

Finally, we consider a TLC-PCM configuration with three different {\DCache}s, sized at 3\%, 6\%, and 12\% of the dataset. Figure~\ref{fig:tlc_pcm} demonstrates that TLC-PCM can only satisfy the performance target with the largest possible {\DCache}, which brings the overall memory hierarchy's cost back in line with the baseline DRAM+{\DCache}~system. 
Given its marginal improvement in performance/cost, as well as TLC's inherently worse endurance~\cite{stanisavljevic:demonstration}, we conclude that TLC-PCM is unable to act as a viable main memory technology for server applications.
That conclusion is reinforced by the clear superiority of MLC-based alternatives.

In summary, we used our performance/cost model to conduct a case study on four currently offered PCM configurations with different cell densities. 
We showed that the configuration that stores 2 bits per cell (MLC) drastically improves performance/cost of the memory hierarchy by 1.7$\times$, whereas the configurations that store one and three bits per cell (SLC and TLC) are not plausible building blocks for server workloads.
Although the real costs of certain memory devices may vary with time, our design exploration methodology still applies, as it relies on fundamental connections among architectural parameters.
If the costs of certain technologies change, our model still provides the performance/cost tradeoffs for the memory devices in question.
We elaborate on the implications of potential cost changes in \S\ref{sec:disc}.

\input{floats/float-tlc-pcm}

\subsection{3D stacked vs. planar DRAM caches}
\label{sec:planar_stacked_comparison}

In this section, we specifically show the superiority of a {\DCache} over a planar DRAM alternative as the first level in an SCM memory hierarchy. 
For this experiment, we use MLC-PCM as the high-capacity tier, and compare 3D stacked DRAM (\DCache) and planar DRAM caches as the high-performance tier. 
Both cache alternatives are organized with 1KB cache blocks.

Figure~\ref{fig:planar_cache_comp} shows that using a  {\DCache}, sized at 3\% of the backing SCM device, improves application performance by 31\% on average (max 81\% for Data Serving) when compared to a single-level DRAM-only configuration. 
This boost in performance is due to the \DCache's ample internal bandwidth and bank-level parallelism. 
A similarly sized planar DRAM cache fails to meet our performance target of being within 10\% of the single-level DRAM-only system for Data Serving and Web Search.

We choose to comment on these two workloads specifically as they represent the cases with the highest memory bandwidth pressure.
This pressure is particularly pronounced in a two-tier SCM hierarchy, as it becomes amplified by data movement between the cache tier and backing SCM.
The additional evict/fill traffic leads to increased pressure in the memory controller queues of the DRAM cache, and the resulting elevated latencies degrade application performance by up to 16\%.
The high degree of internal parallelism on 3D stacked caches alleviates this increased pressure.

A four-fold increase of the planar DRAM cache's capacity (i.e., to 12\% of the backing SCM capacity) improves performance by 5\%, but has the drawback of diminishing the planar cache's cost advantage over a {\DCache}. 
As a result, a system with a {\DCache} of a modest (3\%) size not only outperforms its alternative with a larger (12\%) planar DRAM cache by 33\% on average but also delivers 16\% better performance/cost. 
We expect the performance/cost difference between 3D stacked and planar DRAM caches to grow in the future, as {\DCache} solutions pave their way from being exotic HPC products \cite{sodani:knights} to large-scale deployments in hyperscale datacenters~\cite{google:tpu2,intel:stratix10}.

\input{floats/float-planar_cache_comparison}

%% file: floats/float-model-eval.tex
\begin{figure}[t]

\centering

        \includegraphics[width=\columnwidth]{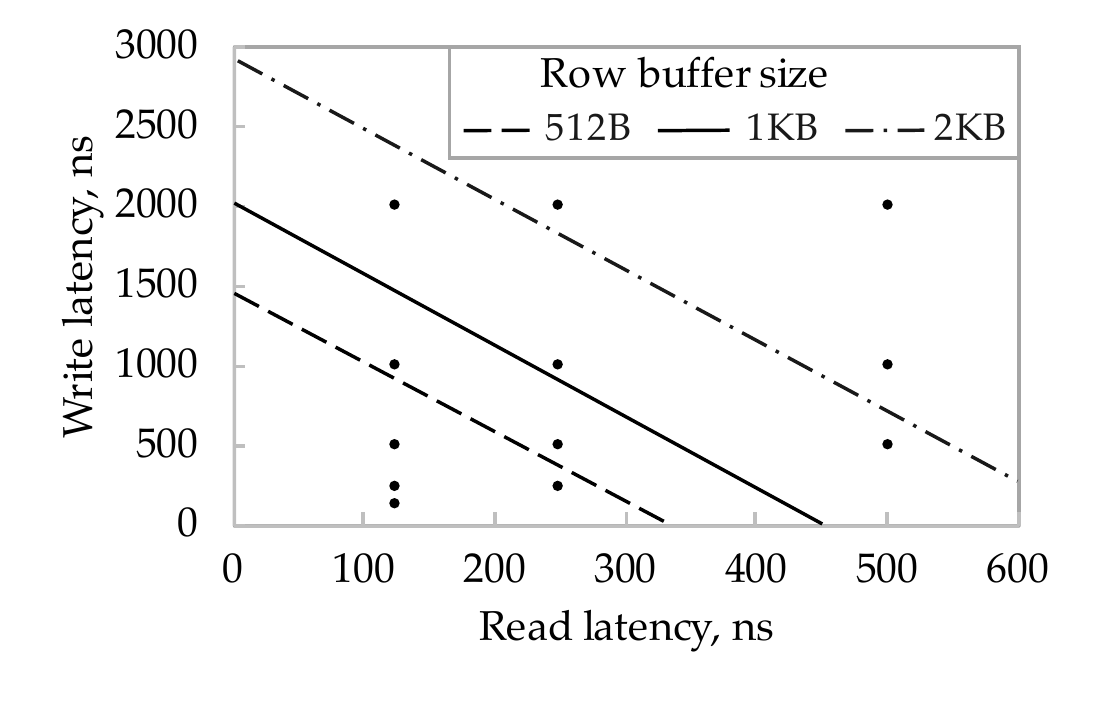}
\caption{Performance model for SCM design space evaluation (planar view of design space frustum from above).}
\label{fig:model_2d}
\end{figure}

%% file: floats/float-model-pcm.tex
\begin{figure}
\centering

        \includegraphics[width=\columnwidth]{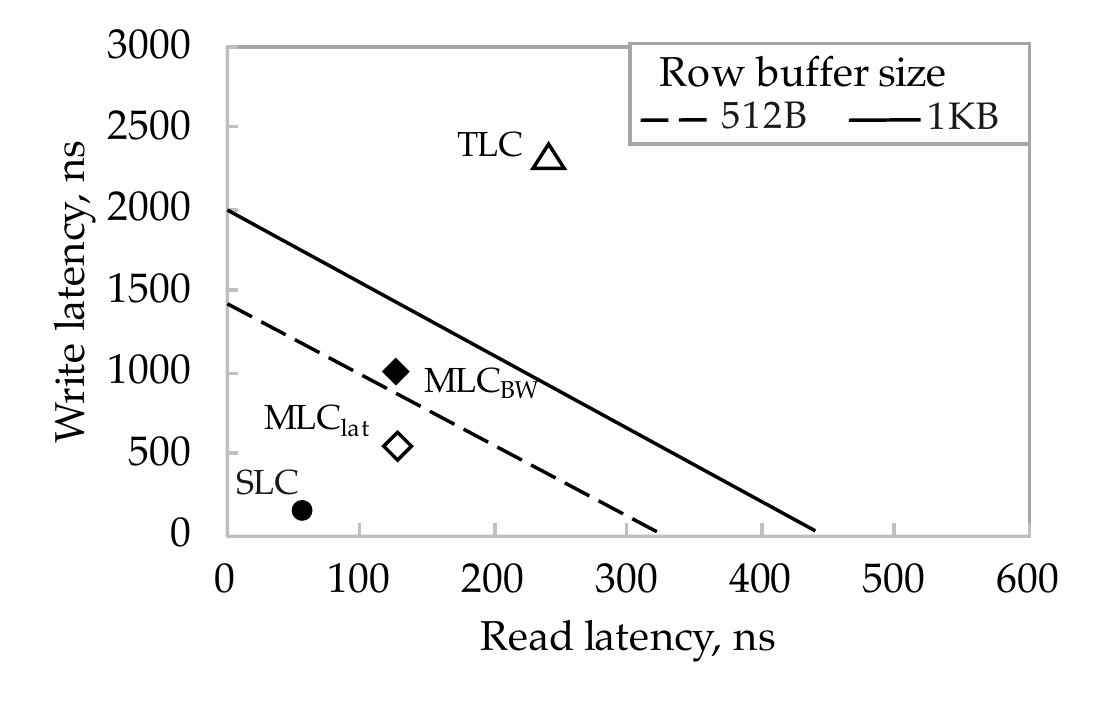}
\caption{Performance model for PCM case study (planar view of design space frustum from above).}
\label{fig:pcm_model2d}
\end{figure}

%% file: floats/float-pcm-res-tbl.tex
\newcolumntype{P}[1]{>{\centering\arraybackslash}p{#1}}

\begin{table}
\begin{center}
\begin{footnotesize}
\def\arraystretch{1.2}	
\begin{tabular}{ |>{\centering\arraybackslash} p{26mm} | P{10mm} | P{6mm} | P{16mm} | P{6mm} |}
\hline
\multirow{2}{*}{Cell configuration} & Perf. \mbox{geomean} & Cache cost & Total memory cost, \% & Perf./ cost \\
\hline
Planar DRAM                   & 1.00      & 0.00  & 1.00  & 1.00   \\
\hline
\DCache (3\%) + DRAM          & 1.31      & 0.22  & 1.22  & 1.07   \\
\hline
\DCache (3\%) + SLC           & 1.30      & 0.22  & 1.22  & 1.06   \\
\hline
\DCache (3\%) + $MLC_{lat}$   & 1.28      & 0.22  & 0.72  & 1.78   \\
\hline
\DCache (3\%) + $MLC_{BW}$    & 1.24      & 0.22  & 0.72  & 1.72    \\
\hline
\DCache (12\%) + TLC         & 1.30      & 0.88  & 1.13  & 1.15    \\

\hline

\end{tabular}
\caption{Performance and cost of various memory hierarchies relative to planar DRAM.\\
}
\label{table:pcm_res}
\end{footnotesize}
\end{center}
\end{table}

%% file: floats/float-tlc-pcm.tex
\begin{figure}[t]
\centering
        \includegraphics[width=\columnwidth]{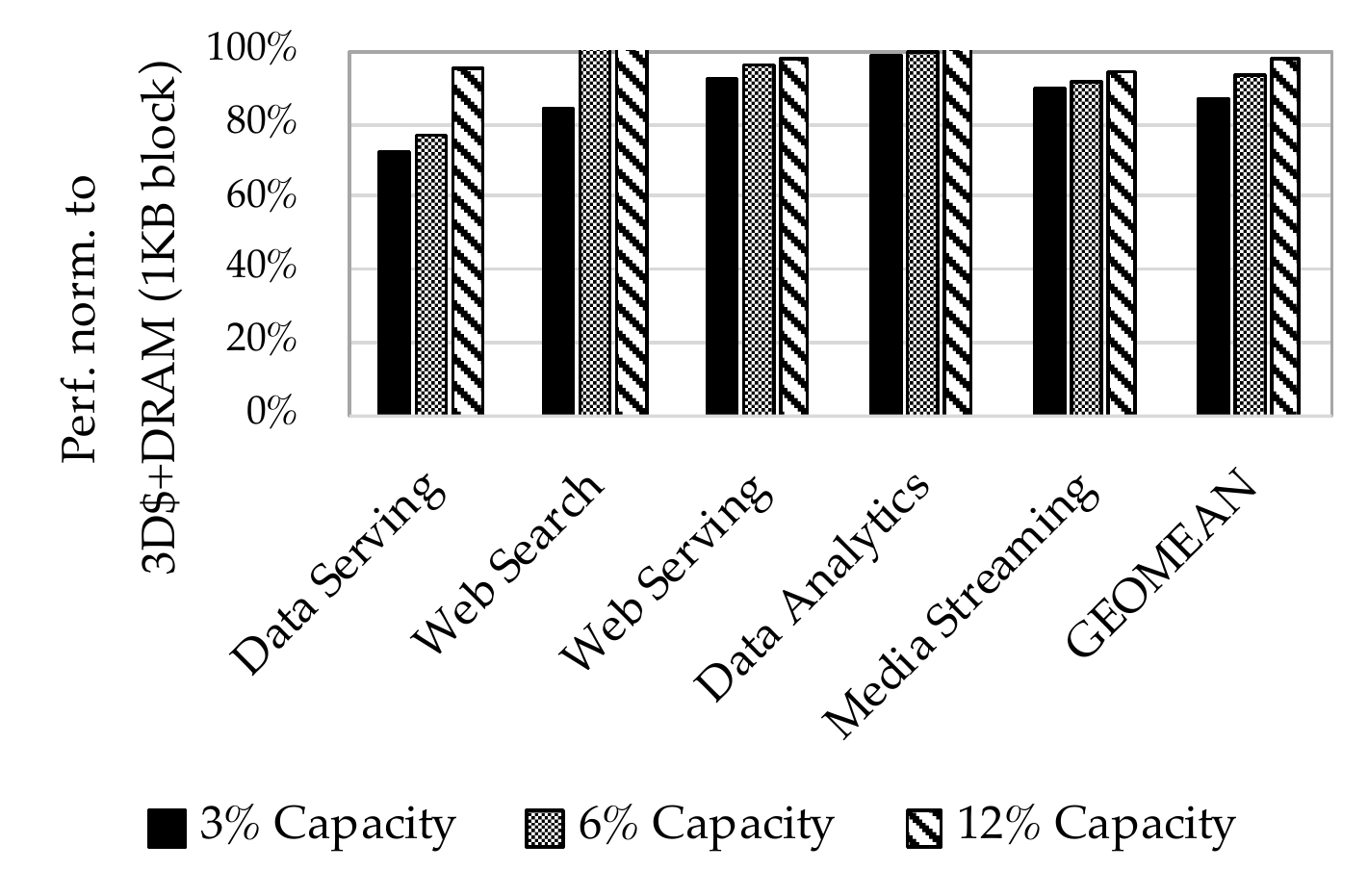}
        \label{fig:block_size}
\caption{TLC-PCM performance with {\DCache}s of different sizes (SCM: $t_{RCD}$=250ns, $t_{WR}$=2350ns as per Table \ref{table:pcm}).
}
\label{fig:tlc_pcm}
\end{figure}

%% file: floats/float-planar_cache_comparison.tex
\begin{figure}[t]
\centering
        \includegraphics[width=\columnwidth]{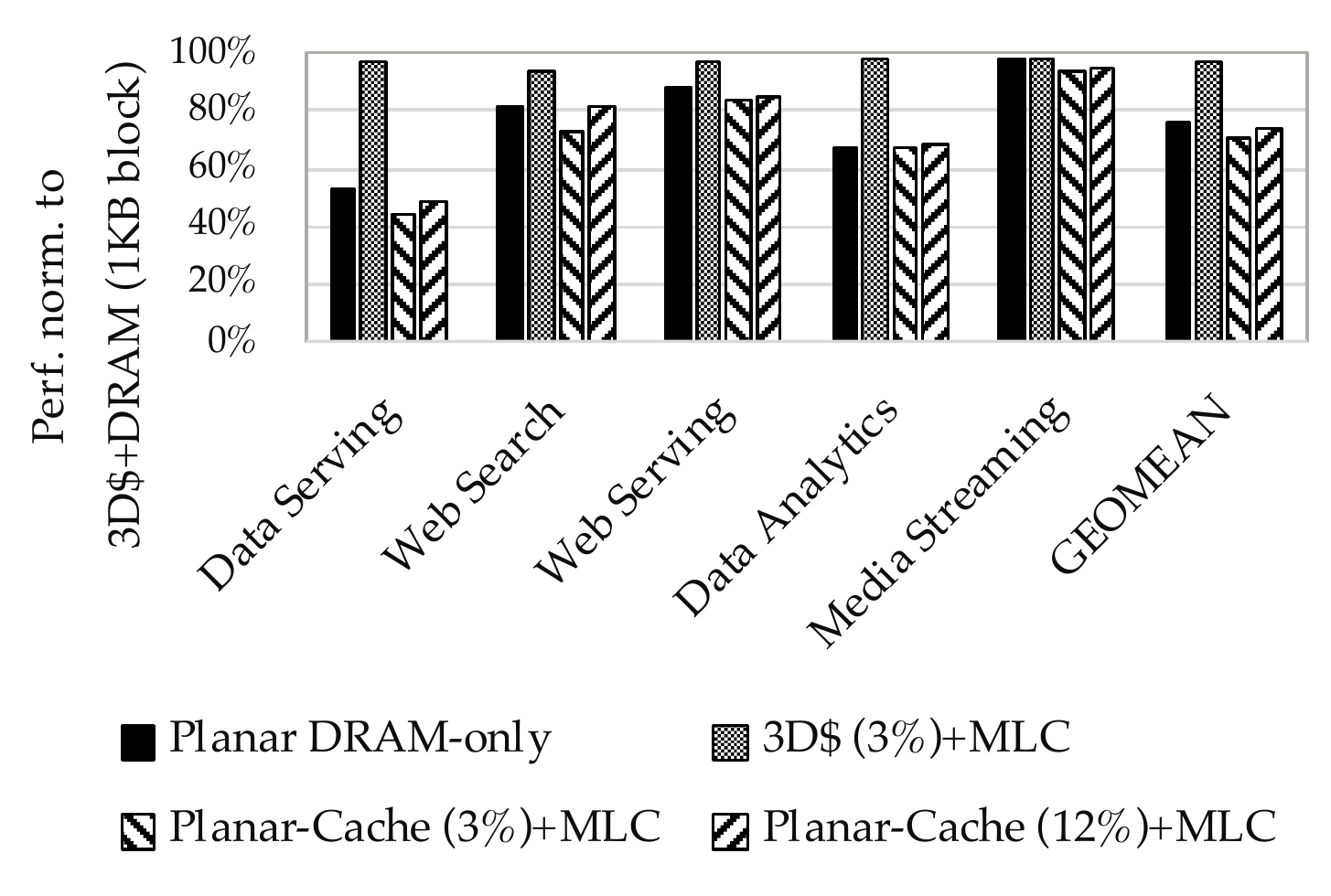}
        \label{fig:planar_cache_comparison}
\caption{MLC-PCM performance with 3D stacked and planar DRAM caches. (MLC-PCM: $t_{RCD}$=250ns, $t_{WR}$=2350ns).}
\label{fig:planar_cache_comp}
\end{figure}

%% file: discussion.tex
\section{Discussion}
\label{sec:disc}

\noindent\textbf{Sensitivity to SCM and 3D stacked DRAM cost.}
Industry has already started to adopt 3D stacked DRAM solutions at a large scale, including but not limited to AMD and NVIDIA GPUs~\cite{anandtech:nvidia,pcgamer:what}, Intel/Altera and Xilinx FPGAs~\cite{wissolik:virtex,intel:stratix10}, and emerging AI solutions like Google's TPU and Wave Computing's DPU~\cite{jouppi:in-datacenter,google:tpu2,wave:dpu}; in higher volumes, the cost of 3D stacked technology is expected to drop.

Cheaper 3D stacked caches will improve the cost-effectiveness of our {\DCache}+SCM hierarchy, mainly by reducing absolute cost rather than motivating the deployment of larger caches, as we show that they have diminishing performance returns (Figures \ref{fig:locality} and \ref{fig:tlc_pcm}). 
Our conclusions regarding the cache block and row buffer sizes required to amortize high SCM latencies only rely on workloads exhibiting spatial locality, and remain unaffected by cost. Significant DRAM cache cost reductions could affect assumptions related to our case studies; e.g., an equivalently priced cache with 4$\times$ capacity could make TLC-PCM technologies viable.\\

\noindent\textbf{SCM persistence aspects.} In this work, we investigated building cost-efficient memory hierarchies for in-memory services, by leveraging emerging  high-density SCM technologies. We demonstrated that SCM hierarchies can approach near-DRAM speed, by amortizing high SCM latencies with bulk memory accesses. While we ignored the additional qualitative benefit of persistence, future architectures featuring SCM will likely also leverage the persistence feature for attaining lower-cost durability \cite{balmau:flodb,nalli:analysis,liu:dudetm,zhang:mojim,zhang:study}. 

To make persistent memory updates durable, the software has to explicitly, and usually synchronously, flush cache lines from the volatile cache hierarchy, which may lead to severe performance degradation. 
In \S\ref{sec:background}, we demonstrated that fine-grain accesses severely degrade SCM's internal bandwidth. 
Using our latency amortization insights, performance-critical software should strive for bulk accesses, which can naturally achieved by using log-based software systems. 
For example, DudeTM~\cite{liu:dudetm} separates performance-critical threads which run user transactions and generate logs, from background threads that apply updates in-place. 
As a transactional log entry usually spans a few kilobytes, intelligently written software should be able to amortize the latency cost of writing logs to SCM.

Alternatively, our small {\DCache} can also be made persistent using lithium-ion batteries already available in modern OpenCompute racks~\cite{ocp:battery_std}. Microsoft practitioners have already demonstrated the maturity of this technology~\cite{dragojevic:no_compromises,kateja:viyojit}, and its ability to reliably back up hundreds of gigabytes of DRAM-resident data upon a power failure; this capacity assumption perfectly matches the {\DCache} capacities we consider in this work.
If the {\DCache} is made persistent, then log-generating threads will not need to explicitly use bulk accesses;
their writes can transparently go to the {\DCache} without needing to explicitly ensure the logs have been replicated to the non-volatile SCM.

%% file: related.tex
\section{Related Work}
\label{sec:related}

Our work draws inspiration from extensive studies in the fields of server architecture and memory systems.
In this section, we look at the relationship between our work and prior proposals.\\

\noindent\textbf{DRAM caches for servers.}
Previous studies have leveraged the wide high speed interface and highly parallel internal structure of {\DCache}s to mitigate the "memory bandwidth wall" found in server applications \cite{volos:fat}.
Block-based organizations~\cite{huang:atcache,loh:efficiently,qureshi:fundamental,sodani:knights,ayers:memory,chou:batman,chou:bear,young:dice,chou:cameo} tend to perform better in the presence of temporal locality, while page-based ones~\cite{jiang:chop,lee:fully,volos:fat} favor applications with spatial locality.
Scale-out workloads tend to possess more spatial locality than temporal~\cite{volos:bump}, motivating the use of page-based caches~\cite{volos:fat}. However, increasing core counts in servers introduce bandwidth concerns as well, rendering simple page-based designs that overfetch data suboptimal. 
The Footprint~\cite{jevdjic:die-stacked} and Unison~\cite{jevdjic:unison} caches mitigate this overfetch problem, by leveraging an access pattern footprint predictor, at the cost of slightly increasing the DRAM cache's miss ratio. 
Our work extends these observations to SCM hierarchies, and shows that {\DCache}s in our context should also be page-based, since transferring data in large chunks amortizes the long latency of accessing SCM. 
The increased cost of DRAM cache misses in our context precludes the addition of Footprint and Unison's predictor mechanism.
Specifically, Footprint cache's slightly increased miss ratio offsets its lower SCM bandwidth utilization resulting in virtually the same performance as the page-based DRAM cache design, primarily because of its miss traffic's fine granularity (64B).

Volos et al.~\cite{volos:fat} also propose a hierarchical memory system for servers, featuring a \DCache~backed by planar DRAM. Their findings indicate that the \DCache~should be sized to host 10--15\% of the dataset, which is 3--5$\times$ larger than the \DCache~we advocate. 
However, their design goals are different, as they scale down the frequency of the memory bus to DDR-1066 to save energy. 
In our work, we assume a commodity DDR4 interface, since SCM DIMMs (adhering to the NVDIMM-P standard) are expected to be DDR4-compliant~\cite{snia:nvdimm}. 
The median data rate of DDR4 is DDR-2666~\cite{micron:ddr4}, which is significantly higher than that used in previous work~\cite{volos:fat}. In our setting, a \DCache~that contains 3\% of an application's dataset is enough to capture most of the available spatial locality (Figure~\ref{fig:block_sensi}) while the backing memory's interface offers enough bandwidth to serve the fraction of traffic that the \DCache~does not filter. That traffic ends up being slightly higher than what is generated when using a 10--15\% {\DCache}~\cite{volos:fat}, but is still well within the SCM's bandwidth capacity.

Other researchers have proposed to mitigate long SCM latencies by using conventional planar DRAM DIMMs for hardware-managed caches~\cite{qureshi:scalable}, OS-based page migration \cite{hirofuchi:raminate} and application-assisted data placement \cite{dulloor:data}.
Applying these designs in the context of server workloads will expose the lack of internal parallelism in planar DRAM devices~\cite{volos:fat}, leading to excess request queuing and therefore inflated latencies.
Qureshi et al.~\cite{qureshi:scalable} proposed using a hardware-managed DRAM cache in front of high-capacity SCM to mitigate its high access latency. Our work extends the state of the art with a thorough analysis of server applications, demonstrating the superiority of a page-based {\DCache} over planar and block-based alternatives, and proposing a methodology to help memory architects design the most cost-effective SCM solutions.\\

\noindent\textbf{DRAM and SCM flat integration.} 
While we considered a two-level hierarchy with a hardware-managed DRAM cache as our baseline (\S\ref{sec:workloads}), a number of prior proposals consider an alternative memory system organization: flat integration of SCM and DRAM on a shared memory bus~\cite{agarwal:thermostat,dulloor:data,hirofuchi:raminate}.
In these proposals, software is responsible for placing the data on the heterogeneous DIMMs, relying on heuristics to optimize for performance~\cite{agarwal:thermostat, dulloor:data} or energy efficiency~\cite{hirofuchi:raminate}.
The major strength of this organization is its compatibility with the existing DDR4 interface; however, this also turns out to be a key weakness, as it is optimized for fine-grain (64B) data accesses rather than bulk transfers which are preferred by SCM. 

We find that preserving a unified memory interface and using it for heterogeneous memories has two important shortcomings. 
First, having a unified DDR interface between heterogeneous memories fundamentally limits the headroom for technology-specific optimizations. 
Specifically, we showed that careful selection of the data transfer granularity (i.e., prioritizing bulk accesses) is essential to mitigate SCM's much higher access latencies.

Second, because of the expected order-of-magnitude capacity mismatch between SCM and DRAM, it is likely that a workload's hot dataset fraction won't fit exclusively in DRAM. 
For example, Agarwal et al.~\cite{agarwal:thermostat} find that avoiding significant performance degradation (i.e., $\textless$3--10\%) requires severely limiting demand memory traffic going to the SCM (down to 30--60MB/s) that in turn requires a large fraction (50--70\%) of an application's dataset to be DRAM-resident. Consequently, (i) applications will likely suffer from a shortage of DRAM as well as low utilization of the vast SCM capacity, and (ii) the part of the hot dataset that will end up being directly served from the SCM will be accessed at a fine granularity (64B), giving up the opportunity for SCM latency amortization via coarse-grained accesses.
In contrast, our proposal of deploying SCM with an appropriately sized page-based \DCache~will deliver high performance even when the applications' hot dataset fraction cannot entirely fit in the available DRAM capacity.


Overall, we consider hardware and software mechanisms to be complementary to each other: hardware can provide low latency access for direct demand traffic to the SCM, whereas software can efficiently optimize data movement across memory channels taking advantage of different high-level characteristics (non-volatility, low static power, high endurance, etc.) of the heterogeneous DIMMs.\\

\noindent\textbf{SCM device optimizations.}
Since SCM write bandwidth is heavily constrained by current limitations inside the DIMM, industry prototypes have limited-sized row buffers~\cite{suzuki:nonvolatile}.
In order to reduce peak write power, prior work uses a technique called \emph{differential writes}, that detects the subset of bits that actually change their values during a write restoration, which are often as few as 10-15\%~\cite{lee:architecting,qureshi:scalable}.
This technique shrinks the effective write current and enables greater row buffer sizes, which is critical to our techniques in this paper. 
Fine-grained power management techniques at the DIMM level have a similar goal but operate above the circuit and cell level~\cite{hay:preventing,jiang:fpb}, and mainly focus on manipulating the limited power budget.

To reduce SCM DIMM latency through the use of SRAM row buffers, Yoon et al. proposed a row buffer locality aware caching policy for heterogeneous memory systems \cite{yoon:row}, allocating addresses that cause frequent row buffer misses in DRAM.
Lee et al. proposed architecting SCM row buffers as small associative structures, decoupled from data sensing, to reduce row buffer conflicts and leverage temporal locality as this design allows for several simultaneously open rows~\cite{lee:architecting}. 
However, server workloads exhibit poor temporal but abundant spatial locality~\cite{volos:bump}. \\

\noindent\textbf{Tackling the SCM read/write disparity.}
As most SCM technologies show significant disparities in read and write latencies~\cite{lu:tutorial,wong:stanford,xu:understanding}, prior work has proposed various mechanisms to mitigate the effects of slow SCM writes. 
At the application level, researchers have proposed new algorithms that generate less write traffic~\cite{chen:rethinking,viglas:write}. 
At the hardware level, Fedorov et al. augmented a conventional LRU eviction policy to reduce the eviction rate of written data~\cite{fedorov:ari}.
To mitigate head-of-line blocking of critical reads behind long latency iterative SCM writes, prior work has proposed enhanced request scheduling mechanisms, which cancel or delay writes and allow reads to bypass them~\cite{arjomand:boosting,qureshi:improving,zhao:firm}.
Qureshi et al. proposed a reconfigurable SCM hierarchy that is able to dynamically change its mode of operation between high performance and high capacity~\cite{qureshi:morphable}. 
At the device level, Wang et al. suggested buffering writes separately from the SCM row buffers to move data array writes off the critical path~\cite{wang:building}.
Finally, Sampson et al. proposed using fewer write iterations to improve SCM access latencies at the cost of data precision~\cite{sampson:approximate}.

We group this diverse list of prior work together because our work obviates the need for any special hardware extensions related to read/write latency disparity. Our design methodology helps system designers determine the range of tolerable read/write latency pairs based on the target application domain's spatial locality characteristics and SCM device's row buffer size.
Furthermore, our insights show that SCM designers can sacrifice device speed to improve other non-performance characteristics. 
For example, Mellow Writes~\cite{zhang:mellow} shows that slowing down writes can increase the lifetime of ReRAM by orders of magnitude, while Zhang et al. demonstrate a similar tradeoff for PCM~\cite{zhang:balancing}.
When considering whether or not to adopt such a technique, our performance model provides concrete evidence to architects that extended latencies can indeed be tolerated given the opportunity to amortize them with large row buffers.

%% file: conclusion.tex
\section{Conclusion}
\label{sec:conclusion}

The arrival of emerging storage-class memory technologies has the potential to revolutionize datacenter economics, allowing online service providers to deploy servers with far greater capacities at decreased costs.
However, directly using SCM as an alternative for DRAM raises significant challenges for server architects, as its higher activation latencies are unacceptable for datacenter applications with strict response time constraints.
We show that although fully replacing DRAM with SCM is not possible due to increases in memory access latency,
a carefully architected 3D stacked DRAM cache placed in front of the SCM allows the server to match the performance of a state-of-the-art DRAM-based system.
The abundant spatial locality present in server applications favors a page based cache organization, which enables amortization of long SCM access latencies.

As SCMs come in a plethora of densities and performance grades, we provide a methodology that helps construct a performance model for a given set of applications, prune the broad design space, and design the most cost efficient memory hierarchy combining a modestly sized 3D stacked DRAM cache with the SCM technology of choice.
We demonstrate the utility of our methodology by performing a case study on a number of phase-change memory devices and show that 2-bit cells currently represent the only cost-effective solution for servers.

%% file: ack.tex
\section*{Acknowledgements}
\label{sec:ack}

The authors thank the anonymous reviewers for their invaluable feedback and insightful comments, as well as Steve Byan, Frederic T. Chong, Mario Drumond, James Larus, Virendra J. Marathe, Arash Pourhabibi, Yuan Xie, and the members of the PARSA and DCSL groups at EPFL for their support and numerous fruitful discussions.

This work has been partially funded by the Nano-Tera \textit{YINS} project, Huawei Technologies, the Swiss National Science Foundation project 20021\_165749, CHIST-ERA \textit{DIVIDEND}, and the European Commission's H2020 \textit{Eurolab-4-HPC} project.
